\documentclass[preprint,12pt]{elsarticle}

\usepackage{graphicx}
\usepackage{amssymb}
\usepackage{lineno}
\usepackage{array}
\usepackage{color}
\usepackage{adjustbox}
\usepackage{multirow,hhline}
\usepackage{xfrac}
\usepackage{kotex}
\usepackage{gensymb}

\journal{Physics of Fluids}

\begin{document}
\begin{frontmatter}

\title{Numerical Study on Interactional Aerodynamics of a Quadcopter in Hover with Overset Mesh in OpenFOAM}

\author[label1]{Young Min Park}
\author[label1]{Solkeun Jee\corref{mycorrespondingauthor}}
\cortext[mycorrespondingauthor]{Corresponding author}
\ead{sjee@gist.ac.kr}
\address[label1]{School of Mechanical Engineering, Gwangju Institute of Science and Technology, 123 Cheomdan-gwagi-ro, Buk-gu, Gwangju 61005, South Korea}

\begin{abstract}
Interactional aerodynamics of a quadcopter in hover is numerically investigated in this study.
The main objective is to understand major flow structures associated with unsteady airloads on multirotor aircraft.
The overset mesh approach is used to resolve flow structures in unsteady simulation using the flow solver OpenFOAM.
The current computational study demonstrates that aerodynamic interaction between quadcopter components strongly affects the rotor wake, generating interesting vortical structures. 
Multiple rotors in close proximity generate $\Omega$-shaped vortical structures merged from rotor-tip vortices. 
The fuselage of the current quadcopter deflects the wake flow of the four rotors towards the center of the vehicle.
Such interactional aerodynamics, i.e., rotor-rotor and rotor-fuselage interaction, varies the inflow condition of a rotor blade during the rotor revolution.
Therefore, the quadcopter experiences unsteady airloads per rotor revolution.
Our study indicates that a typical quadcopter would experience 8/rev thrust variations, which are a combined outcome from 4/rev thrust variations on the rotor and 2/rev fluctuations on the fuselage.
The current understanding of interactional aerodynamics could help to design reliable and efficient multicopter aircraft.
\end{abstract}


\end{frontmatter}


\section{Introduction}
The aim of this study is to investigate interactional aerodynamics of multirotor aircraft, especially focusing on aerodynamic interaction between multiple rotors, and between a rotor and a fuselage.
The close proximity of rotors and fuselage in multicopter configurations leads to complex aerodynamic interaction, which generates unsteady airloads. 
Unsteady aerodynamic loads can reduce the overall performance of a vehicle \cite{Throneberry2021, Russell2018, Yoon2017A, VenturaDiaz2018}, potentially cause vibration and sectional loading in rotors \cite{Intaratep2016, Zawodny2020, Sheen2022, Wu2022}, and degrade the control authority \cite{Hoffmann2011}.
With increasing demand for more compact configurations in recent multirotor aircraft, stronger interactional aerodynamics is anticipated among distributed rotors and an airframe.
Therefore, it is imperative to understand interactional aerodynamics of a multicopter in order to reduce aircraft operation risk and optimize aircraft performance.

Experimental studies have been conducted to characterize the aerodynamic performance of various multicopter vehicles in a hovering state \cite{Russell2018, Intaratep2016}.
These studies show that a multicopter produces relatively lower thrust compared to its scaled counterpart of an isolated single rotor.
Russell et al. \cite{Russell2018} conducted comprehensive wind tunnel tests and showed that the aerodynamic performance of multicopter is degraded compared to its isolated rotor counterpart. 
Intaratep et al. \cite{Intaratep2016} also found that the vehicle thrust is lower than the linear progression as the number of rotors increases in multicopter configurations. 
Therefore, it is intuitive to conjecture that the overall performance of a multicopter would be associated with interactional aerodynamics among rotors and fuselage.

The effect of rotor-rotor interaction has been investigated through experimental \cite{Ning2018, Lee2021} and computational \cite{Lee2020, Alvarez2020} studies, revealing that the rotor-rotor interaction affects the rotor wake around a multicopter with decreasing an effective angle of attack and the overall vehicle thrust. 
Ning et al. \cite{Ning2018} performed experiments with a twin-rotor configuration and measured less thrust with higher thrust fluctuation as the distance between the two rotors decreases. 
Ning et al. also observed that rotor-rotor interaction distorts tip vortices, resulting in a different inflow condition toward the rotor blade compared to the isolated single rotor. 
Lee et al. \cite{Lee2021} examined the impact of rotor-rotor interaction on a quadrotor configuration and found that the rotor-rotor interaction deflects the rotor wake towards an adjacent rotor. 
The presence of tip vortices from the adjacent rotor increases the magnitude of the axial induced velocity, leading to a reduced effective angle of attack and the loss of the sectional thrust.
Rotor interactional effects have been computationally investigated by Alvarez et al. \cite{Alvarez2020} and Lee et al. \cite{Lee2020} with twin and quad rotor configurations, respectively. 
Both computational studies predicted a reduced rotor thrust, consistent with previous multirotor experiments. 
Alvarez et al. \cite{Alvarez2020} visualized merged vortices from blade tips between rotors.
Lee et al. \cite{Lee2020} showed that reduction in the rotor thrust is caused by the decreased effective angle of attack when the rotor blades are closely located. 
Rotor-rotor interaction induces an upwash flow between rotors and distorts the rotor wake towards the center of the rotors. 
Similar rotor-rotor interaction has also been reported in a side-by-side rotor configuration by Sagaga et al. \cite{Sagaga2021}. An $\Omega$-shaped vortical structure is formed by the rotor-rotor interaction, which leads to variations of sectional thrust on the rotor blade.

Rotor-fuselage interaction is another critical aspect that affects multicopter aerodynamics. 
Especially, flow phenomena affected by the fuselage of a multirotor are often different from those of a traditional single main-rotor vehicle. 
Potsdam et al. \cite{Potsdam2005} investigated tiltrotor configurations in hovering flight and observed flow recirculation between rotors, caused by the fuselage. 
The recirculating flow is ingested into the rotor, which reduces the rotor performance. 
In multicopter configurations with more than two rotors, Yoon and Ventura Diaz \cite{Yoon2017A, VenturaDiaz2018} showed that the majority of the reduction in the thrust comes from an airframe being placed in the rotor wake. 
The fuselage also modifies major vortical structures which generate unsteady airloads \cite{Yoon2017A, VenturaDiaz2018}.

Despite aforementioned studies, interactional aerodynamics of a multicopter has not been thoroughly investigated.
In particular, it remains unexplored how rotor-rotor and rotor-fuselage interactions generate unsteady vortical structures and fluctuating airloads.
To elucidate potential effects of complex aerodynamic interactions on multirotor aircraft, i.e. reduction of vehicle performance, structural fatigue, and rotor vibration, understanding of interactional aerodynamics of a multicopter is needed.
Therefore, the objective of this study is to investigate the significant influence of interactional aerodynamics on unsteady airloads acting on a multicopter.
Since unsteady airloads are highly related with unsteady flow fields around a multicopter, a detailed study on major vortical structures is also critical, which is pursued in this study as well.
This study will delve further into unsteady airloads, particularly focusing on sectional thrust, which can be potentially contributed to the structural fatigue of rotor blades and fuselage.
Moreover, unsteady airloads and their associated frequencies will be investigated, because they can lead to vehicle vibration.

Unsteady Reynolds-averaged Navier-Stokes (URANS) simulation is employed to investigate the unsteady flows around a quadcopter in this study. 
The URANS approach has been successfully used in previous studies such as large-scale rotorcraft \cite{Yoon2017A, VenturaDiaz2018, Potsdam2005, Sagaga2021, Chaderjian2012, Strawn2002} and small-scale multicopters \cite{Hwang2015, Paz2021, CarrenoRuiz2022, Thai2022}. 
The overset mesh method in the flow solver OpenFOAM \cite{Weller1998ATA, Jasak1996, Katavic2018, Senjanovic2020} is explored for the simulation of a quadcopter here.
An overset mesh method is efficient in resolving multiple rotors simultaneously with a stationary fuselage \cite{Yoon2017A, VenturaDiaz2018, Thai2022, Hwang2015}. 
Yet, the overset mesh method in OpenFOAM framework has not been applied to a multicopter for the best of the authors' knowledge. 
Therefore, the current overset approach in OpenFOAM will be validated with relevant experiments \cite{Russell2018, Ning2018} and computational studies \cite{VenturaDiaz2018} where the well-recognized overset code OVERFLOW was used.

The organization of this paper is as follows. 
Computational methods are provided in section \ref{sec:methods} including a specific quadcopter case in section \ref{sec:quadcopter_flowCondition}, computational grids in section \ref{sec:computationalGrids}, the current flow solver in section \ref{sec:numericalSettings}, and the overset method in section \ref{sec:overset}. 
Computational results are validated in section \ref{sec:valid} for an isolated single rotor (section \ref{sec:single}) and a quadcopter configuration (section \ref{sec:quadcopter}) against relevant data in the literature. 
Interactional aerodynamics of the current quadcopter is discussed in section \ref{sec:unsteady} with in-depth analysis on unsteady flow fields and airloads.
Finally, conclusions are given in section \ref{sec:conclusions}.

\section{Methods} \label{sec:methods}
\subsection{Quadcopter and flow condition} \label{sec:quadcopter_flowCondition}

\begin{figure}[htb!] \centering
	\includegraphics[width=.9\linewidth]{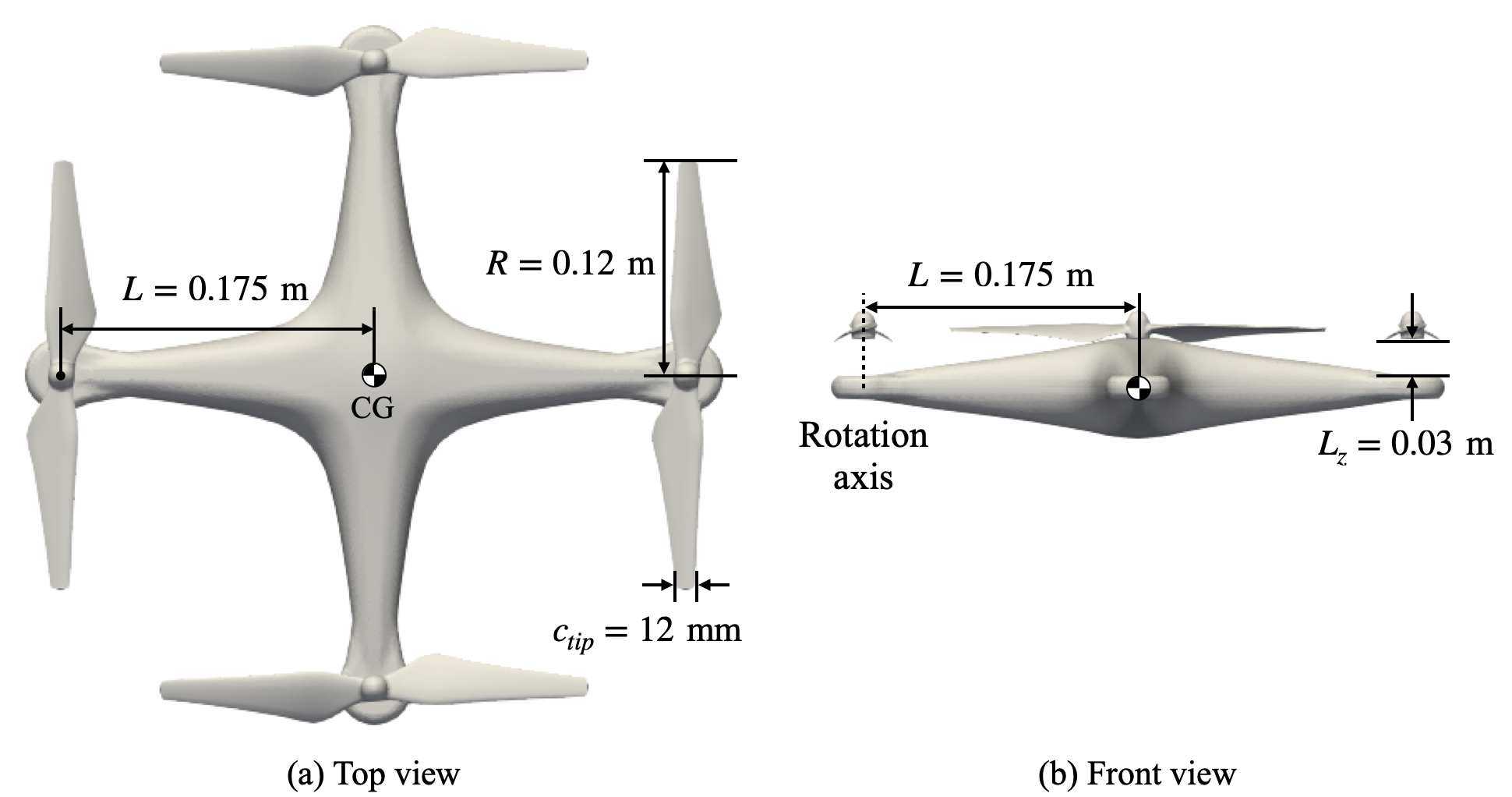} 
	\caption{Geometry of the quadcopter DJI Phantom 3 with the simplified fuselage.}
	\label{fig:DJI_CAD}
\end{figure}

The commercial quadcopter DJI Phantom 3 is numerically investigated in this study. The quadcopter includes four equipped rotors (DJI 9450) and the main portion of the fuselage as shown in Fig. \ref{fig:DJI_CAD}. 
Geometry information for the current quadcopter is obtained from high-resolution 3D scanning with the resolution of 18 $\mu$m. 
Additional components of the fuselage, such as landing gears, battery, and camera are not included here, due to their little influence on the overall aerodynamic performance of the quadcopter \cite{VenturaDiaz2018}.

The radius of the rotor blade is $R$ = 0.12 m and the tip chord length is $c_{tip}$ = 12 mm. 
Each rotor is located on its fuselage arm with the distance $L$ = 0.175 m from the center of the fuselage, and the rotor is slightly offset by $L_z$ = 0.03 m from the fuselage arm because the rotor motor is not resolved in the current study (see Fig. \ref{fig:DJI_CAD}). 
The minimum distance between rotors is about 0.067$R$. The minimum distance between the rotor and the fuselage is 0.108$R$ when the rotor tip is positioned just above the fuselage arm. 

\begin{figure}[htb!] \centering
	\includegraphics[width=\textwidth]{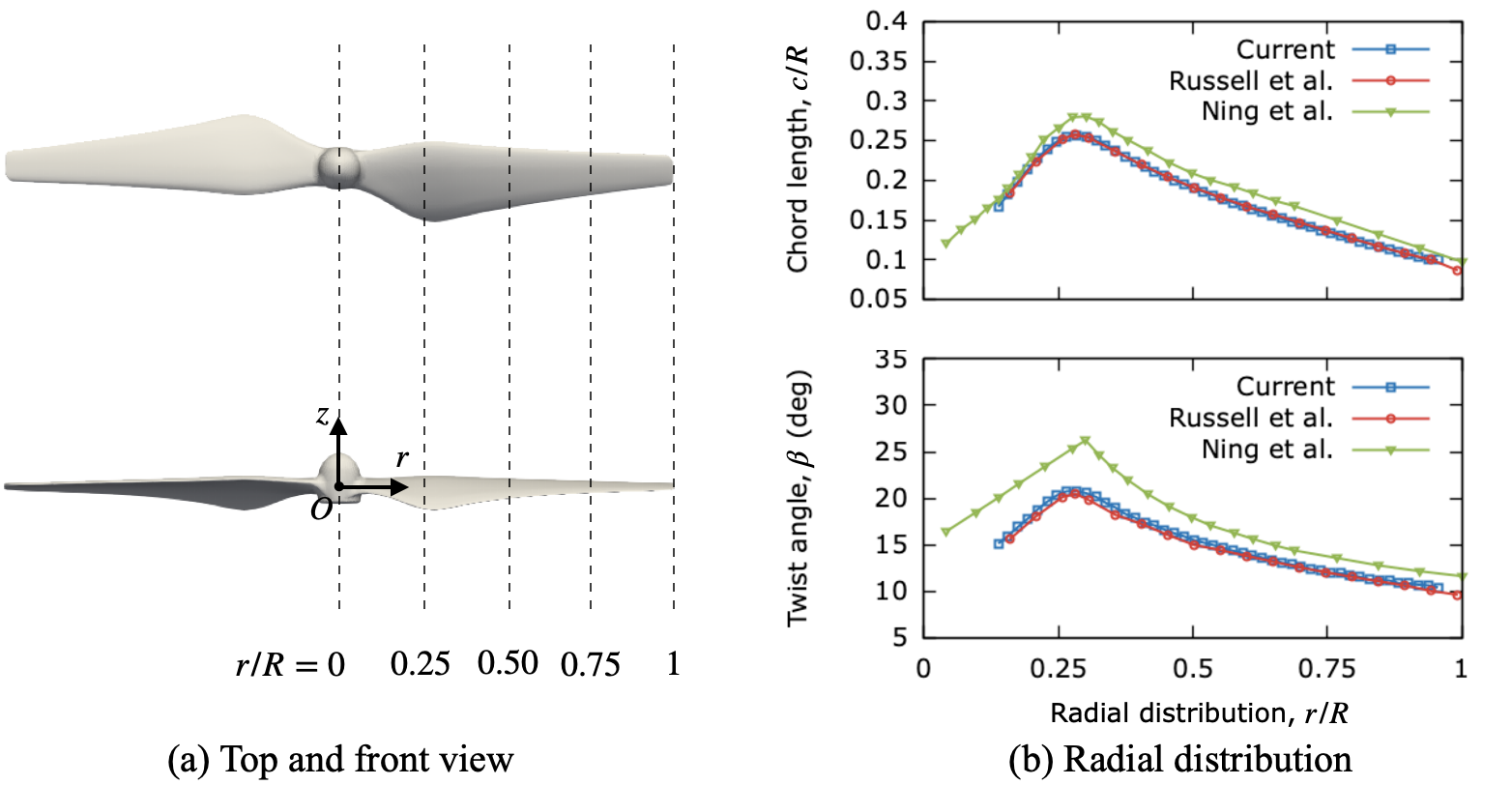}
	\caption{Geometry and radial distributions of the current isolated single rotor with similar rotors from previous experiments \cite{Russell2018, Ning2018}.}
	\label{fig:rotorDistribution}
\end{figure}

The detailed geometry of the isolated single rotor is shown in Fig. \ref{fig:rotorDistribution}. 
The origin of the rotor coordinates $(r, z)$ is positioned at the center of the rotor hub, and the location $z = 0$ aligns with the rotor tip (see Fig. \ref{fig:rotorDistribution}(a)). 
The radial distributions of the blade chord $c$ and the twist angle $\beta$ are shown in Fig. \ref{fig:rotorDistribution}(b), in comparison with similar rotors from previous experiments \cite{Russell2018, Ning2018}.
The subsequent validation section will include a comparative analysis with these experimental data.

\begin{figure}[htb!] \centering
	\includegraphics[width=\textwidth]{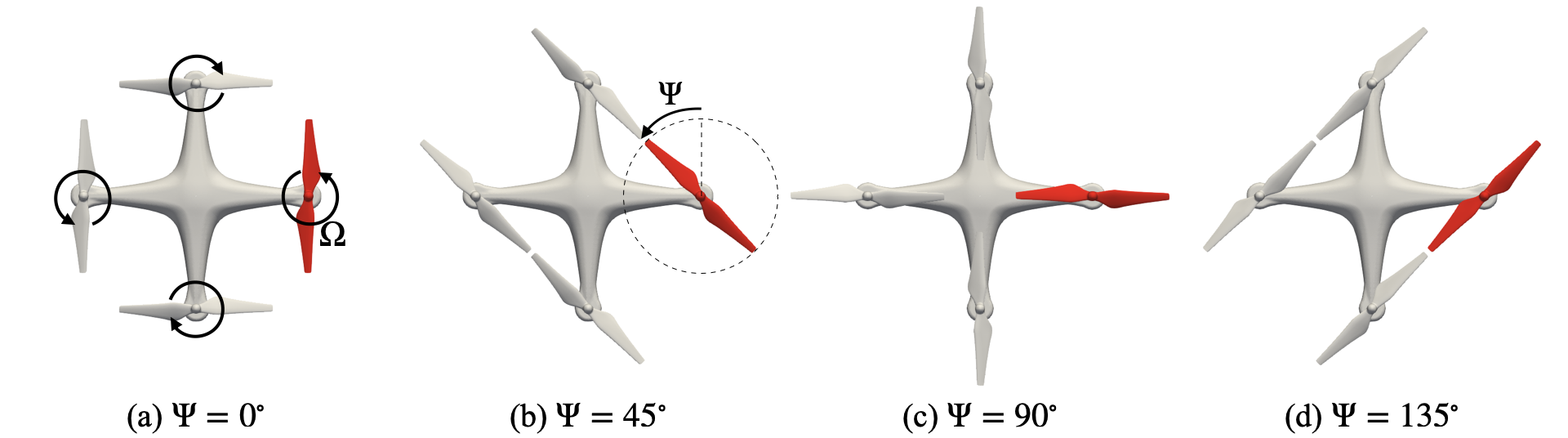}
	\caption{Arrangements of the rotors in the quadcopter during the half rotor revolution.}
	\label{fig:DJI_revolution}
\end{figure}

In this study, two configurations in hovering state are simulated: (1) the isolated single rotor and (2) the quadcopter. Both configurations are simulated with rotor speeds at $\Omega$ = 3000, 3500, 5000 and 7000 RPM following previous experiments \cite{Russell2018, Ning2018}. 
The range of the rotor tip velocity is $U_{tip}$ = 38--88 m/sec, which corresponds to the rotor tip Reynolds number $Re_{tip}$ = 0.4--1.1 $\times 10^5$. 
The air density $\rho$ is set to $\rho$ = 1.2 $\rm{kg/m^3}$ as the same condition from the previous experiment \cite{Russell2018}.
The azimuthal angle of the rotor in the quadcopter configuration is denoted by $\Psi$ for the highlighted rotor shown in Fig. \ref{fig:DJI_revolution}.
Each rotor rotates either in a clockwise or a counter-clockwise direction with a constant RPM. 
It should be noted that during actual operation, the rotor speeds of each rotor can vary over time, resulting in phase modulation.
Nonetheless, such phase modulation is not included in this study to focus on the rotor-rotor and rotor-fuselage interactions for given azimuthal angles.

\subsection{Computational grids} \label{sec:computationalGrids}

The overset mesh approach is used for the numerical treatment of moving rotors. 
Off-body background grids are generated using snappyHexMesh, the grid generation tool in OpenFOAM \cite{Weller1998ATA, Jasak1996}, and near-body grids around rotors are generated with the commercial software Pointwise \cite{Pointwise}. 
Detailed description of computational grids for both the isolated single rotor and the quadcopter is provided below.

\subsubsection{Isolated single rotor}

\begin{figure}[htb!] \centering
	\includegraphics[width=.45\textwidth]{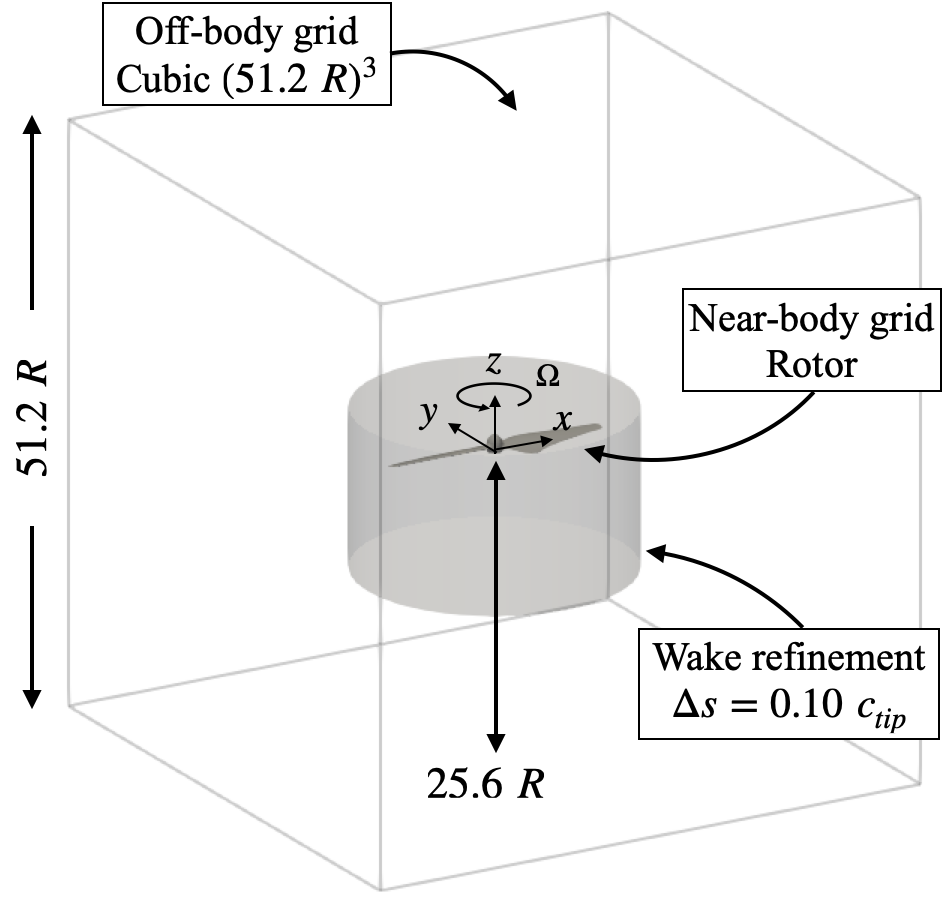}
	\caption{Schematic diagram of the computational domain for the isolated single rotor.}
	\label{fig:gridDomain_single}
\end{figure}

\begin{figure}[htb!] \centering
	\includegraphics[width=\textwidth]{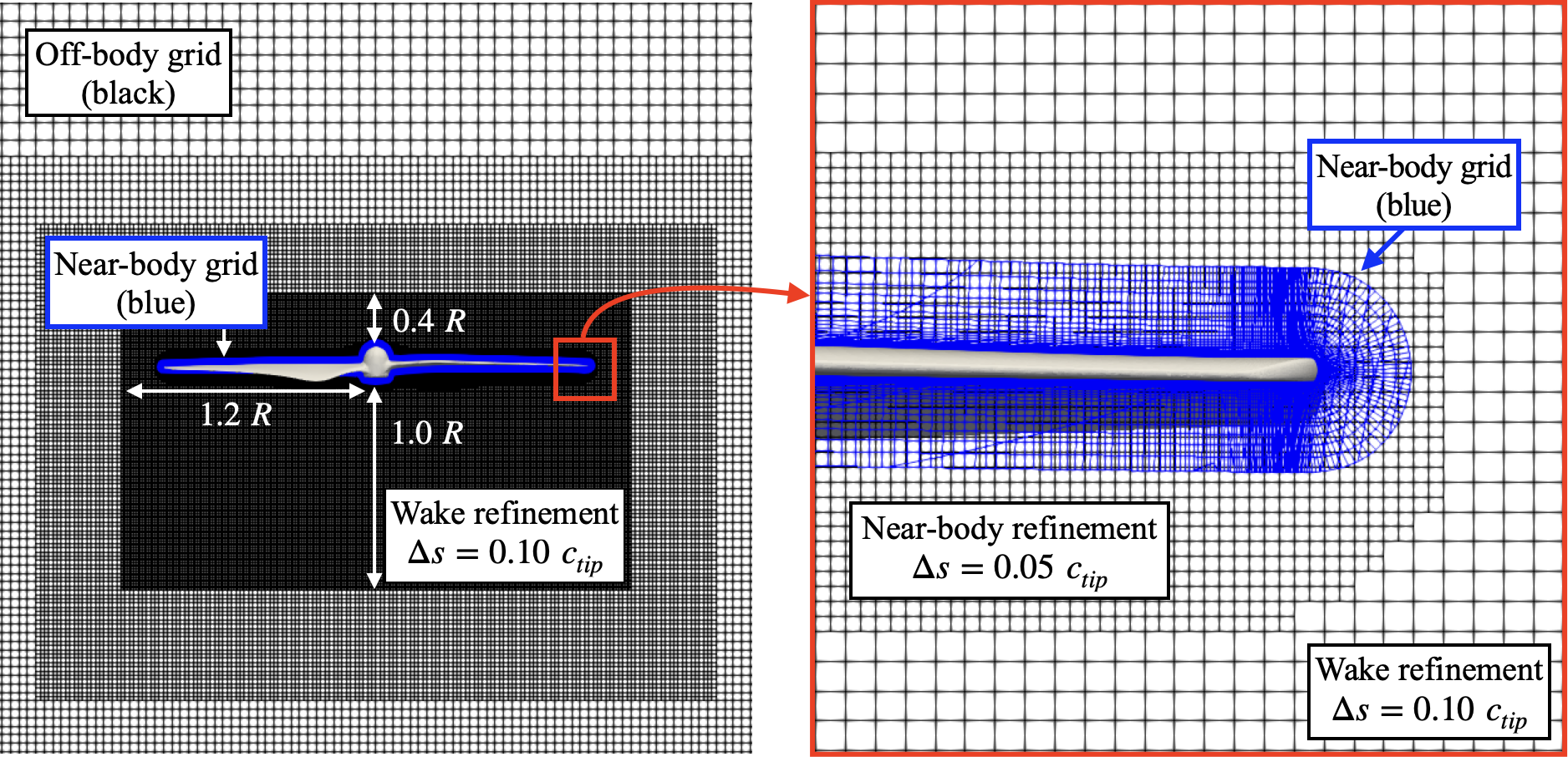}
	\caption{Sectional view of the computational grids around the rotor. The near-body grid is colored as blue, and the off-body grid is colored as black, respectively.}
	\label{fig:gridOB_single}
\end{figure}

Computational grids for the isolated single rotor are composed of a near-body grid around the rotor and an off-body grid which includes the far-field domain as shown in Fig. \ref{fig:gridDomain_single}. 
Structured isotropic grids are used for the overall computational domain $(51.2R)^3$, and the size of the overall computational domain is set to be similar to previous computations of quadcopters \cite{VenturaDiaz2018, Yoon2017A}.
The near-body rotor grid is placed at the center of the off-body grid.
A portion of the off-body grid is refined around the near-body rotor grid, referred to as wake refinement in Fig. \ref{fig:gridDomain_single} and \ref{fig:gridOB_single}. 
This refined region, with a grid size $\Delta s = 0.1c_{tip}$, extends approximately $1R$ into the wake to capture the vortical wake (see wake refinement in Fig. \ref{fig:gridOB_single}), following the recommendation of Chaderjian et al. \cite{Chaderjian2012}.
Additional grid refinement with a grid size of $\Delta s = 0.05c_{tip}$ is used to improve the interpolation between the overlapped grids around the near-body grid (see near-body refinement in Fig. \ref{fig:gridOB_single}).
The off-body grid contains $N_{vol, ob} =$ 13.5 million volume cells, with 5.8 and 4.5 million cells allocated for the wake refinement and near-body refinement regions, respectively.

\begin{figure}[htb!] \centering
	\includegraphics[width=\textwidth]{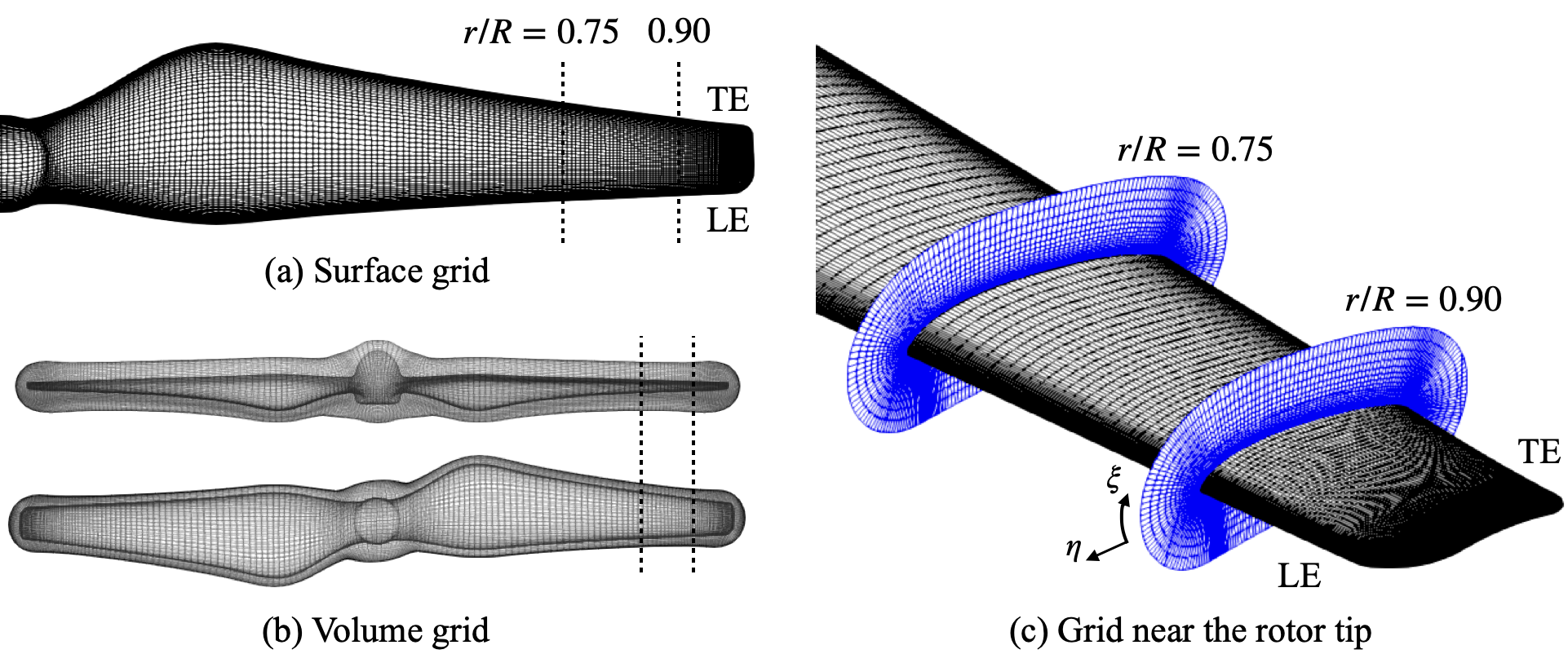}
	\caption{The near-body grid for the rotor used in the simulation.}
\label{fig:gridRotor}
\end{figure}

The near-body rotor grid used in the simulation is shown in Fig. \ref{fig:gridRotor}.
The surface grid is generated with grid sizes near the leading edge (LE) $\Delta s_{LE} = 1.3\times10^{-3}c_{tip}$ and the trailing edge (TE) $\Delta s_{TE} = 4.7\times10^{-4}c_{tip}$, respectively (see Fig. \ref{fig:gridRotor}(a)).
The volume grid is generated by extruding the surface mesh in the wall-normal direction to capture viscous boundary layers with a sufficiently small grid size $\Delta s_\eta = 1.7\times10^{-4}c_{tip}$  (see Fig. \ref{fig:gridRotor}(b)).
This provides the first wall-normal grid size $\Delta s_\eta^+ < 1$ on most of the blade surfaces. 
O-type sectional grids at two specific radial positions, $r/R = $ 0.75 and 0.9, are also shown in Fig. \ref{fig:gridRotor}(c). 
The near-body grid contains $N_{surf, nb} = $ 0.11 million surface cells and $N_{vol, nb} = $ 3.5 million volume cells, respectively.
Thus, the total grid count, including both the near-body and off-body grids, is $N_{tot} = $ 17.0 million.

\subsubsection{Quadcopter} \label{sec:gridQuadcopter}

\begin{figure}[htb!] \centering
	\includegraphics[width=.95\textwidth]{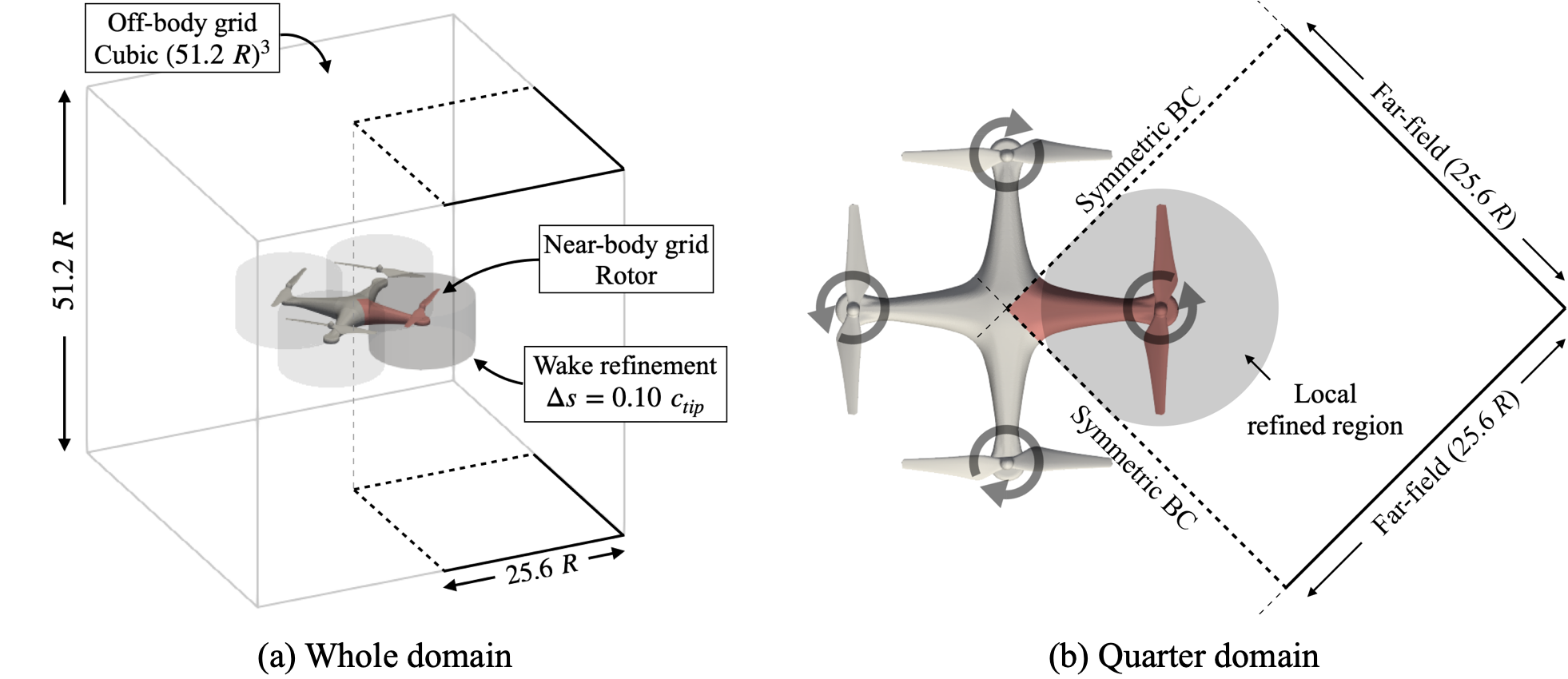} 
	\caption{Schematic diagram of the computational domains for the quadcopter case.}
	\label{fig:gridDomain_quad}
\end{figure}

\begin{figure}[htb!] \centering
	\includegraphics[width=.9\textwidth]{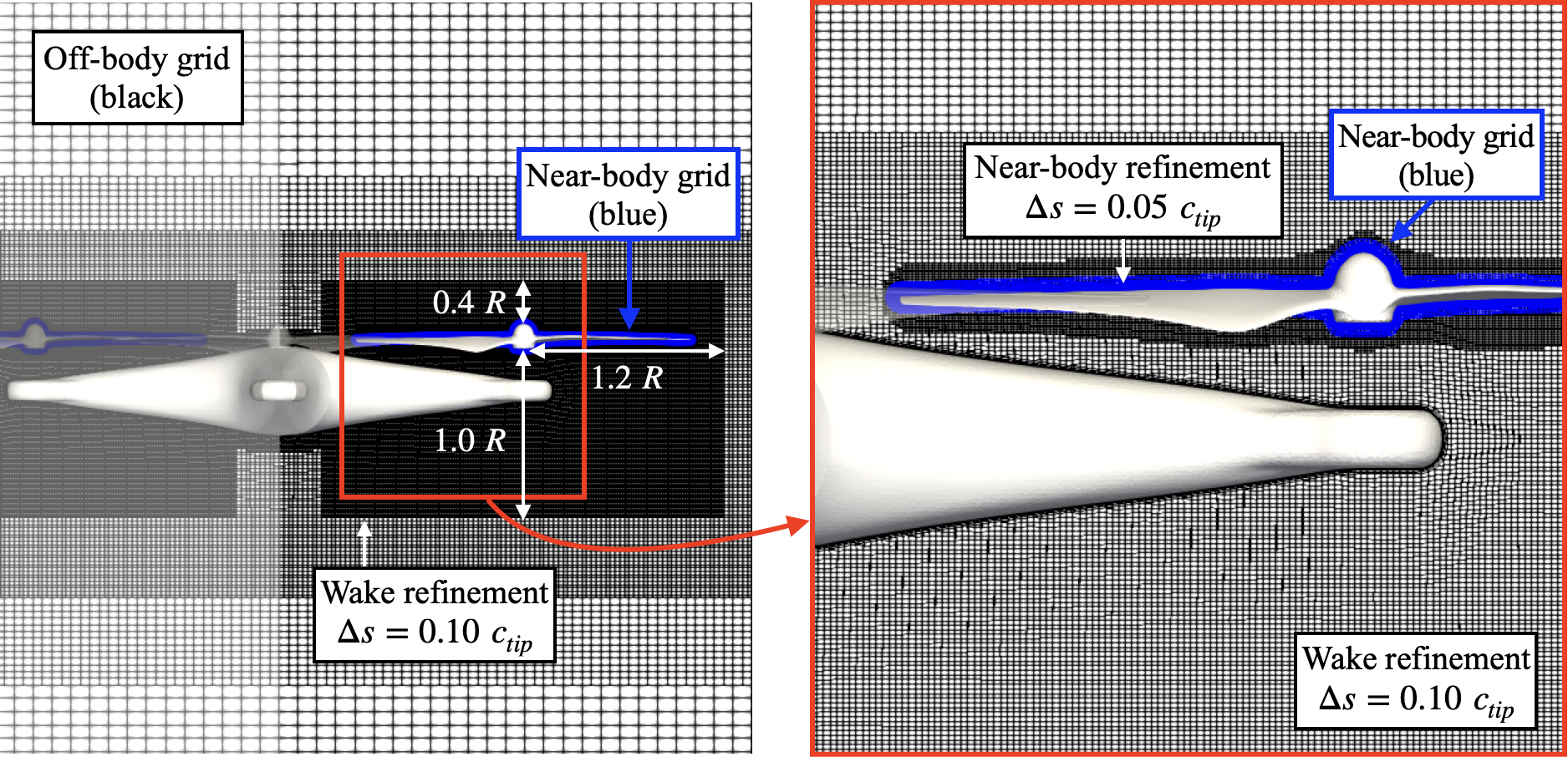} 
	\caption{Sectional view of the computational grids around the quadcopter. Near-body grids is colored as blue, and the off-body grid is colored as black, respectively}
	\label{fig:gridOB_quad}
\end{figure}

Computational grids for the quadcopter are composed of four near-body rotor grids and an off-body grid including the fuselage as shown in Fig. \ref{fig:gridDomain_quad}(a). 
The fuselage is positioned at the center of the off-body domain with its rotors mounted on each arm. 
The size of the off-body cubic domain is $(51.2R)^3$. The computational domain size is set to be similar to previous computations of quadcopters \cite{VenturaDiaz2018, Yoon2017A}.
The off-body domain consists mostly of structured isotropic grids and split-hexahedra grids near the fuselage surface.
The grid size on the fuselage surface is $\Delta s = 0.1~c_{tip}$, and the first wall-normal grid size with $\Delta s^+<1$ is allocated on the fuselage surface.
The same near-body rotor grid is used for both the single rotor and the current quadcopter simulation.
The wake region of the off-body grid is refined around the near-body rotor, following the same method as in the single rotor case, as shown in Fig. \ref{fig:gridOB_quad}. 
The number of volume cells for the off-body grid is $N_{vol, ob} = $ 46.2 million, and the number of the total volume cells, including both the four near-body grids and the off-body grid, is $N_{tot} = $ 60.0 million.

The quadcopter configuration can be simplified into a quarter configuration, due to the symmetry in the rotor direction and the fuselage as shown in Fig. \ref{fig:gridDomain_quad}(b).
This simplification yields two computational domains for the quadcopter case: (a) a whole domain that incorporates all the rotors and the fuselage, and (b) a quarter domain consisting of one rotor and a quarter portion of the fuselage as shown in Fig. \ref{fig:gridDomain_quad}(a) and \ref{fig:gridDomain_quad}(b), respectively.
The off-body grid for the quarter domain is simply a quarter of the off-body grid for the whole domain. 
The symmetric inner boundaries in the quarter domain enable efficient computations.
The off-body grid for the quarter domain comprises 11.5 million volume cells, while the total volume cells for the quarter domain, including one near-body grid, amount to 15.0 million.
The computational grids used in current study are listed in Table \ref{tab:case}.

\begin{table}[htb!] \centering
\caption{Computational grids used in the current computations. $N_{vol, nb}$: the number of volume cells of the near-body grid; $N_{vol, ob}$: the number of volume cells of the off-body grid; $N_{tot}$: the number of total volume cells.}
\vspace{1.5ex}
\begin{tabular}{c||c|c|c}
\hline
\multirow{2}{*}{} & \multirow{2}{*}{Single rotor} & \multicolumn{2}{c}{Quadcopter} \\
\hhline{~~--}
 & & Quarter domain & Whole domain \\
\hline
Near-body & 1 rotor & 1 rotor & 4 rotors \\
\hline
\multirow{2}{*}{Off-body} & Far-field & \sfrac{1}{4} Far-field & Far-field \\
& only & with the \sfrac{1}{4} fuselage & with the fuselage \\
\hline 
$N_{vol,nb}\times10^{-6}$ & 3.45 & 3.45 & 13.8 \\
\hline
$N_{vol,ob}\times10^{-6}$ & 13.5 & 11.5 & 46.2 \\
\hline
$N_{tot}\times10^{-6}$ & 17.0 & 15.0 & 60.0 \\
\hline
\end{tabular}
\label{tab:case}
\end{table}

\subsection{Flow solver and boundary conditions} \label{sec:numericalSettings}

Incompressible unsteady Reynolds-averaged Navier-Stokes (URANS) equations are numerically solved using the open-source flow solver OpenFOAM \cite{Weller1998ATA, Jasak1996}, in particular, the version foam-extend-4.1.
The incompressible URANS equations are given by
\begin{eqnarray}
\frac{\partial u_i}{\partial x_i}
= 0
\label{eqs:cont}
\end{eqnarray}
\begin{eqnarray}
\frac{\partial u_i}{\partial t}
+ u_j \frac{\partial u_i}{\partial x_j}
= (\nu + \nu_t ) \frac{\partial^2 u_i}{\partial x_j \partial x_j}
- \frac{1}{\rho} \frac{\partial p}{\partial x_j}
\label{eqs:mom}
\end{eqnarray}
where $u_i$ is the velocity vector, $\nu$ is the molecular kinematic viscosity, $\nu_t$ is the turbulent kinematic viscosity, and $p$ is the pressure.

The version SA-noft2-R of the Spalart-Allmaras (SA) model \cite{Spalart1994, Aupoix2003, Dacles-Mariani1995} is used here for URANS computations.
The SA-noft2-R model is given by
\begin{eqnarray}
\frac{\partial \tilde{\nu}}{\partial t}
&& + u_j \frac{\partial \tilde{\nu}}{\partial x_j}
= c_{b1} (\tilde{S} + C_{rot} \min (0, S-\Omega)) \tilde{\nu}
- c_{w1} f_w \bigg( \frac{\tilde{\nu}}{d} \bigg)^2\nonumber\\
&& + \frac{1}{\sigma} \bigg[ \frac{\partial}{\partial x_j}
 \bigg( (\nu + \tilde{\nu}) \frac{\partial \tilde{\nu}}{\partial x_j} \bigg)
+c_{b2} \frac{\partial \tilde{\nu}}{\partial x_i} \frac{\partial \tilde{\nu}}{\partial x_i} \bigg]
\label{eqs:SA}
\end{eqnarray}
where the turbulent viscosity $\nu_t$ is computed from the transport variable $\tilde{\nu}$, $\nu_t = \tilde{\nu} f_{\nu1}$.
For further detailed description for the variables and the functions used in Eq. \ref{eqs:SA}, refer Dacles-Mariani et al. \cite{Dacles-Mariani1995} and Spalart and Allmaras \cite{Spalart1994}.
The rotation correction suggested by Dacles-Mariani et al. \cite{Dacles-Mariani1995} is implemented in the flow solver to accurately predict turbulent production near the tip vortices.

The viscous, convective, and pressure gradient terms employ second-order central, second-order upwind, and least-square schemes, respectively.
The pressure-implicit with the splitting of operators (PISO) algorithm is used for the velocity-pressure coupling \cite{Issa1986}.
The implicit three-level second-order scheme is used for time marching.
Simulations are conducted with an initial time step equivalent to $\Delta \theta = 2.5\degree$ for the first 10 rotor revolutions, and then subsequently reduced to $\Delta \theta = 0.125\degree$ for at least 5 consecutive rotor revolutions.
The initial time step $\Delta \theta = 2.5\degree$ is chosen for quick development of the flow field from a non-physical initial condition, here quiescent flow, following previous quadcopter computations \cite{VenturaDiaz2018, Yoon2017A}. 

Boundary conditions for modeling hovering flight in the current computations are listed in Table \ref{tab:bc}.
The subscript $n$ indicates the normal direction relative to a surface.
For the rotor and fuselage surfaces, the no-slip condition for the velocity and the zero gradient condition for pressure are determined.
The far-field condition allows both inflow and outflow, depending on the flow direction. The velocity gradient is set to zero in the boundary-normal direction, following the reference \cite{Greenshields2022} for the far field. The outflow pressure is $p = p_0$, whereas the inflow pressure is $p=p_0-0.5 |u_i|^2$ where $p_0$ is the stagnation pressure.
For the SA equation, the effective inflow condition $\tilde{\nu} /\nu = 3$ suggested by Spalart and Rumsey \cite{Spalart2007} is used in the far-field condition, where $\tilde{\nu}$ is the model variable and $\nu$ is the molecular kinematic viscosity. 

\begin{table}[htb!]
\centering
\caption{Boundary conditions for the current simulation.}
\vspace{1.5ex}
\begin{tabular}{c||c|c}
\hline
 & Rotor and fuselage & Far-field \\
\hline
        \multirow{2}{*}{Velocity} & \multirow{2}{*}{ no-slip wall } & \multirow{2}{*}{$ \partial u_i / \partial x_n$ = 0} \\
& & \\
\hline
        \multirow{2}{*}{Pressure} & \multirow{2}{*}{$ \partial p / \partial x_n = 0 $} & $p=p_0-0.5 |u_i|^2 $ for inflow \\& & $p=p_0$ for outflow \\
\hline
\multirow{2}{*}{Turbulent viscosity} & \multirow{2}{*}{$\tilde{\nu} = 0$} & $\tilde{\nu}/ \nu = 3$ for inflow\\
& & $ \partial \tilde{\nu} / \partial x_n = 0$ for outflow \\
\hline
\end{tabular}
\label{tab:bc}
\end{table} 

\subsection{Overset mesh approach} \label{sec:overset}

\begin{figure}[htb!] \centering
	\includegraphics[width=.95\textwidth]{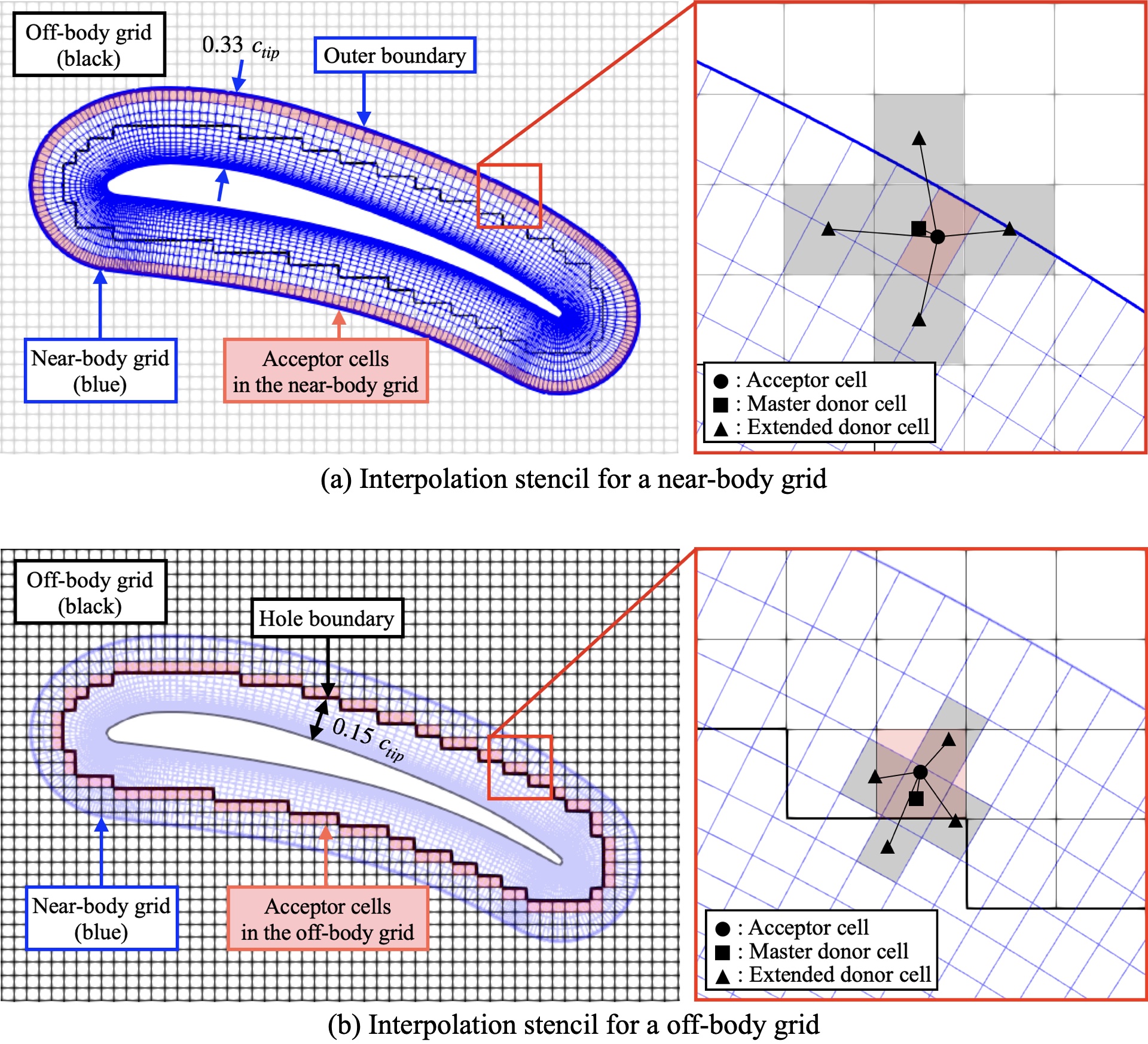}
	\caption{Domain connectivity information for the current overset mesh.}
	\label{fig:overset}
\end{figure}

The overset functionality in foam-extend-4.1 \cite{Jasak1996, Katavic2018} is used to connect the stationary and rotating parts of the computational domain, as shown in Fig. \ref{fig:overset}. 
The current computational domain consists of two grid systems, the near-body grid and the off-body grid. 
Each grid has its own overset boundary: the outer boundary of the near-body grid (see Fig. \ref{fig:overset}(a)) and the hole boundary of the off-body grid (see Fig. \ref{fig:overset}(b)).
Acceptor cells (red cells in Fig. \ref{fig:overset}) are located near the overset boundary of a grid system. 
Flow variables in an acceptor cell are determined by interpolating from a stencil of donor cells in the neighbor grid (grey cells in Fig. \ref{fig:overset}) using inverse distance weighting.
The stencil consists of the nearest donor cell (the master donor cell) and its adjacent cells (the extended donor cells).
The outer boundary of the near-body grid (see Fig. \ref{fig:overset}(a)) is located about $0.33~c_{tip}$ from the blade surface, and the hole boundary of the off-body grid (see Fig. \ref{fig:overset}(b)) is offset roughly $0.15~c_{tip}$ from the blade surface.
The gap between the outer and hole boundaries is set to at least two off-body cells, ensuring that the hole boundary is away from the viscous rotor surface and preventing a situation where an acceptor cannot find donor stencils on the neighboring grid.

\begin{figure}[htb!] \centering
	\includegraphics[width=0.85\textwidth]{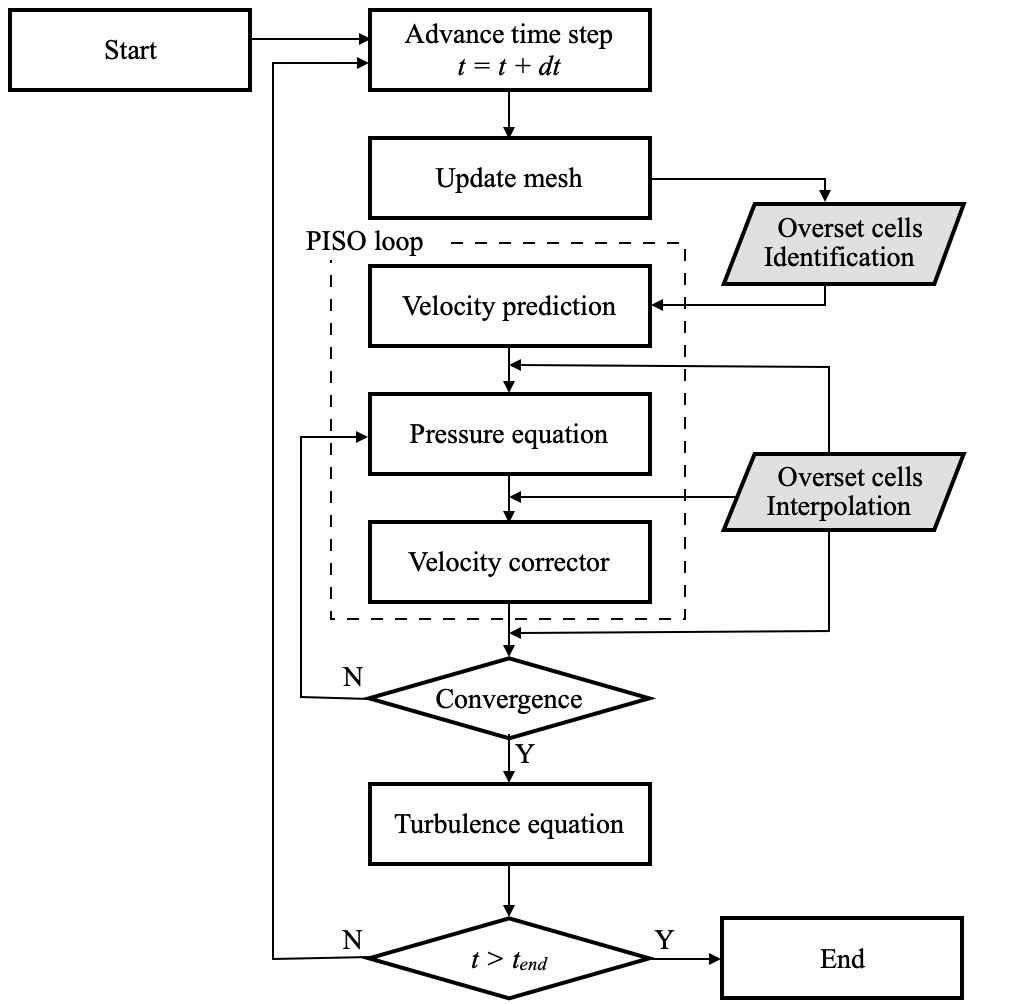}
	\caption{Overset algorithm implemented in OpenFOAM foam-extend-4.1.}
	\label{fig:algorithm}
\end{figure}

In this study, the PISO-based dynamic mesh solver is used including the overset functionality in foam-extend-4.1 \cite{Jasak1996, Katavic2018}.
The algorithm for the current overset solver is depicted in Fig. \ref{fig:algorithm}.
The overset mesh functionality is integrated into procedures for updating meshes and the PISO algorithm (see grey blocks in Fig. \ref{fig:algorithm}). 
At each time step, overset cells and overset boundaries are identified between the rotating rotor mesh and the background mesh.
Based on the domain connectivity information, interpolation between overset cells is carried out in the PISO algorithm.
The PISO loop continues until the residual for flow variables drops below $\mathcal{O}(-8)$.

\section{Validation \label{sec:valid}}

In Section \ref{sec:valid}, the overset mesh method along with OpenFOAM is validated first in two cases: the isolated single rotor (Section \ref{sec:single}) and the quadcopter (Section \ref{sec:quadcopter}). In Section \ref{sec:unsteady}, unsteady aerodynamics on rotors and the fuselage will be investigated.

\subsection{Isolated single rotor \label{sec:single}} 

\begin{figure}[htb!] \centering
	\includegraphics[width=0.55\textwidth]{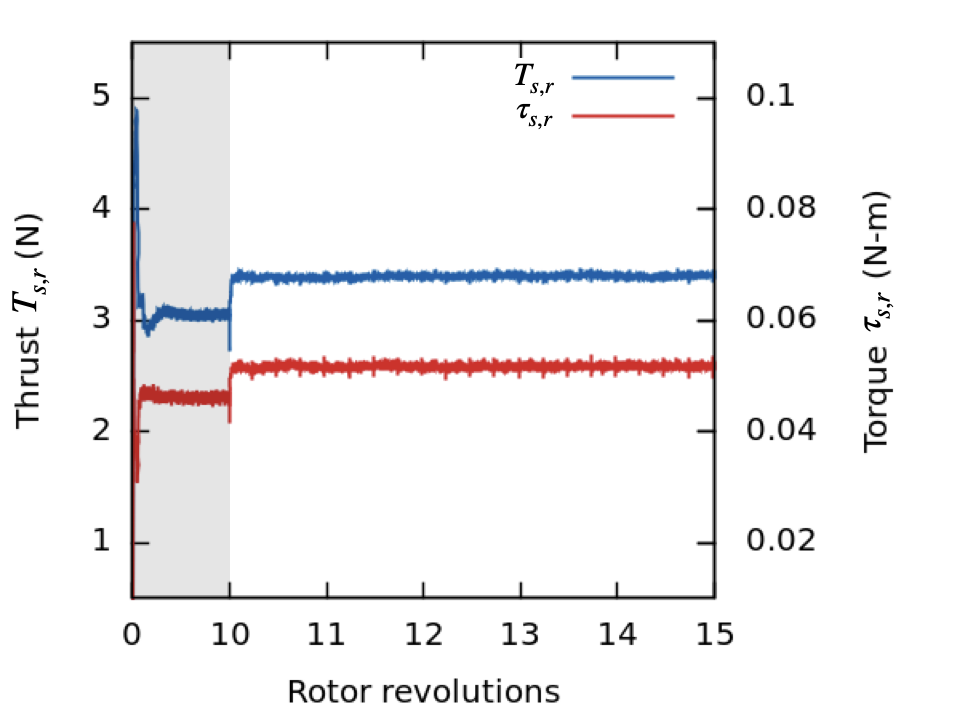}
	\caption{Revolution convergence of the isolated single rotor thrust $T_{s,r}$ and torque $\tau_{s,r}$ at the rotor speed $\Omega$ = 5000 RPM.}
	\label{fig:convergence_single}
\end{figure}

A quasi-steady state of the thrust $T_{s,r}$ and torque $\tau_{s,r}$ for an isolated single rotor $(s,r)$ is achieved in the current unsteady computations as shown in Fig. \ref{fig:convergence_single}. Initial ten rotor revolutions are simulated with a relatively large time step equivalent to $\Delta \theta = 2.5\degree$ for rapid flow development (see the grey region in Fig. \ref{fig:convergence_single}).
After the initial ten rotor revolutions, the time step is reduced to $\Delta \theta = 0.125\degree$.
The rotor thrust and torque experience some initial fluctuation due to the time step change, but reach a quasi-steady state after 11 revolutions.
The differences in the time-averaged thrust and torque between the last two revolutions are only 0.03\%. Therefore, the time-averaged thrust and torque are obtained in the 13th and 14th revolutions.

\begin{figure}[htb!] \centering
	\includegraphics[width=.80\textwidth]{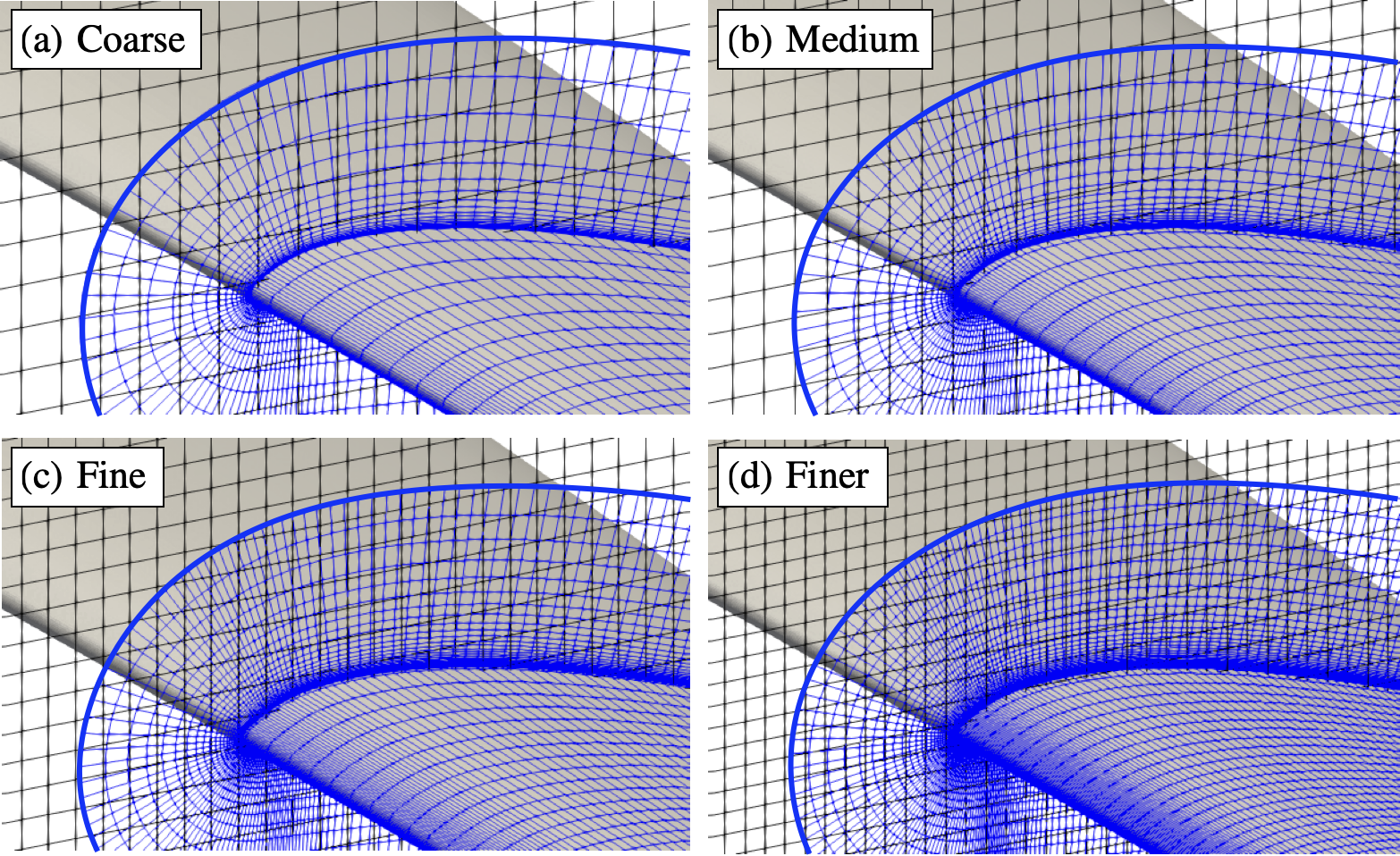}
	\caption{Sectional view of four resolutions of the computational grids. Near-body grid is colored as blue, and the off-body grid is colored as black, respectively.}
	\label{fig:grids}
\end{figure}

\begin{figure}[htb!] \centering
	\includegraphics[width=.80\textwidth]{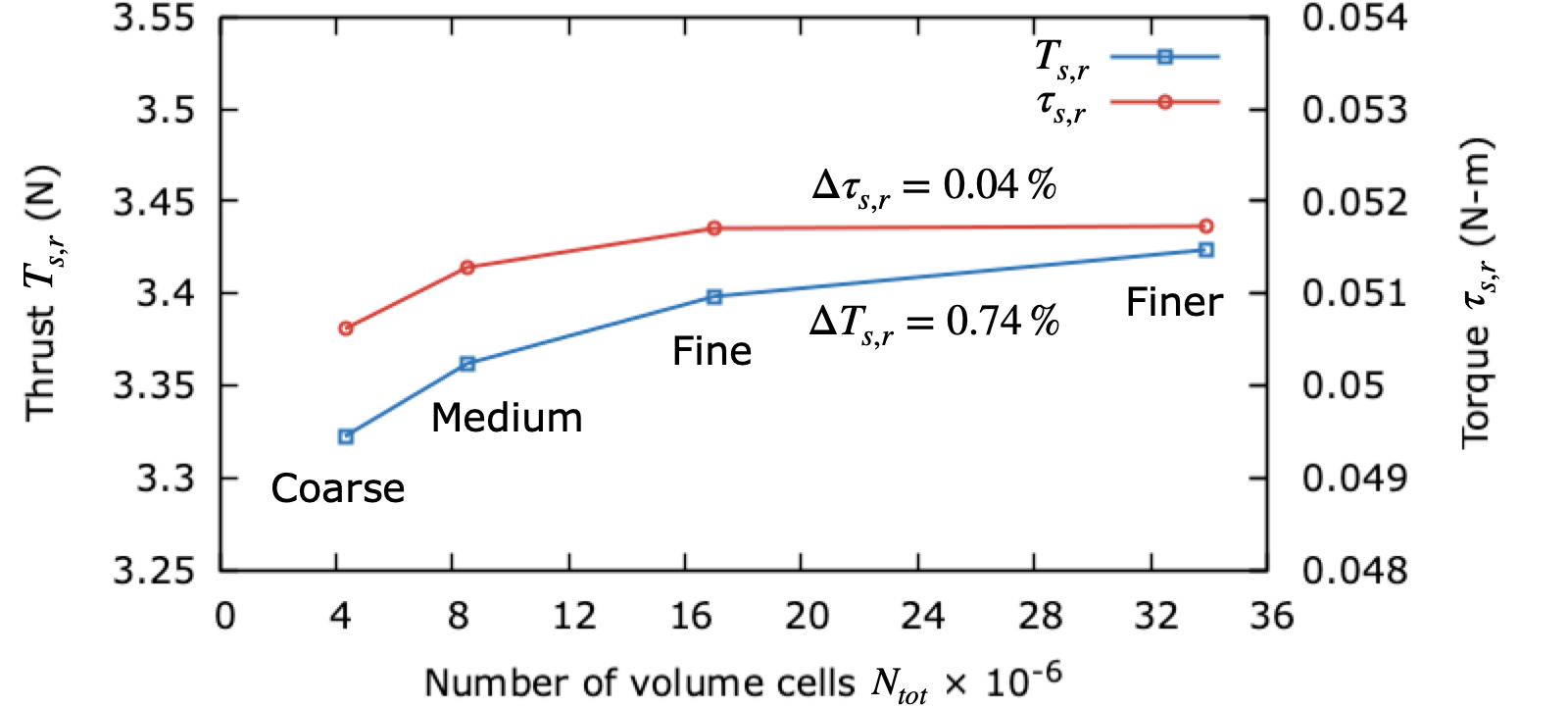}
	\caption{The thrust $T_{s,r}$ and torque $\tau_{s,r}$ for the isolated single rotor with four grid resolutions at the rotor speed $\Omega$ = 5000 RPM.}
	\label{fig:gridConvergence}
\end{figure}

A grid convergence test is conducted for the isolated single rotor using the four grid resolutions.
Based on the current grid (Fine), three additional grid levels (Coarse, Medium, and Finer) of the spatial resolution are designed as listed in Table \ref{tab:grids}. 
A finer grid contains approximately twice the number of grid cells compared to the one-level coarser grid, yielding the grid size is $\sqrt[3]{2}$-times smaller in all the three directions for both the near-body and off-body grid (see Fig. \ref{fig:grids}).
The resulting time-averaged rotor thrust $T_{s,r}$ and torque $\tau_{s,r}$ are shown for each grid system in Fig. \ref{fig:gridConvergence}. 
As the number of volume cells increases, $T_{s,r}$ and $\tau_{s,r}$ converge towards those in the grid Finer.
The grids Fine and Finer provide almost the same thrust and torque with only differences $\Delta T_{s,r}= 0.7\%$ and $\Delta \tau_{s,r}=0.04\%$ between the grids.

\begin{table}[htb!]
\centering
\caption{Four levels of the spatial resolution of grids for the isolated single rotor. $N_{\xi}$: the number of points around the blade in the flow direction; $N_{\eta}$: the number of points in the wall-normal direction; $N_{surf}$: the number of surface cells; $N_{vol}$: the number of volume cells; $N_{tot}$: the number of total volume cells; $\Delta s$: cell size}
\vspace{1.5ex}
\begin{tabular}{c|c||c|c|c|c}
\hline
\multicolumn{2}{c||}{Spatial resolution} 				& Coarse 		& Medium 	& Fine 	& Finer \\ 
\hline
\multirow{4}{*}{Near-body} & $N_{\xi}$				& 160 		& 204		& 258 	& 324 \\ 
					& $N_{\eta}$				& 21 			& 26 			& 32 		& 41 \\ 
					& $N_{surf}$ $\times 10^{-6}$ 	& 0.044 		& 0.071 		& 0.11 	& 0.17 \\
					& $N_{vol}$ $\times 10^{-6}$ 	& 0.88 		& 1.77 		& 3.45 	& 7.01 \\ 
\hline
\multirow{3}{*}{Off-body} 	& $\Delta s/c_{tip}$ near rotor	& 0.079		& 0.063 		& 0.05	& 0.039 \\ 
					& $\Delta s/c_{tip}$ in rotor wake& 0.16		& 0.13		& 0.1 	& 0.079 \\
					& $N_{vol}$ $\times 10^{-6}$	& 3.43 		& 6.74 		& 13.5 	& 26.9 \\ 
\hline
Combined 			& $N_{tot}$ $\times 10^{-6}$ 	& 4.31 		& 8.51 		& 17.0 	& 33.9 \\ 
\hline
\end{tabular}
\label{tab:grids}
\end{table}

\begin{figure}[htb!] \centering
	\includegraphics[width=\textwidth]{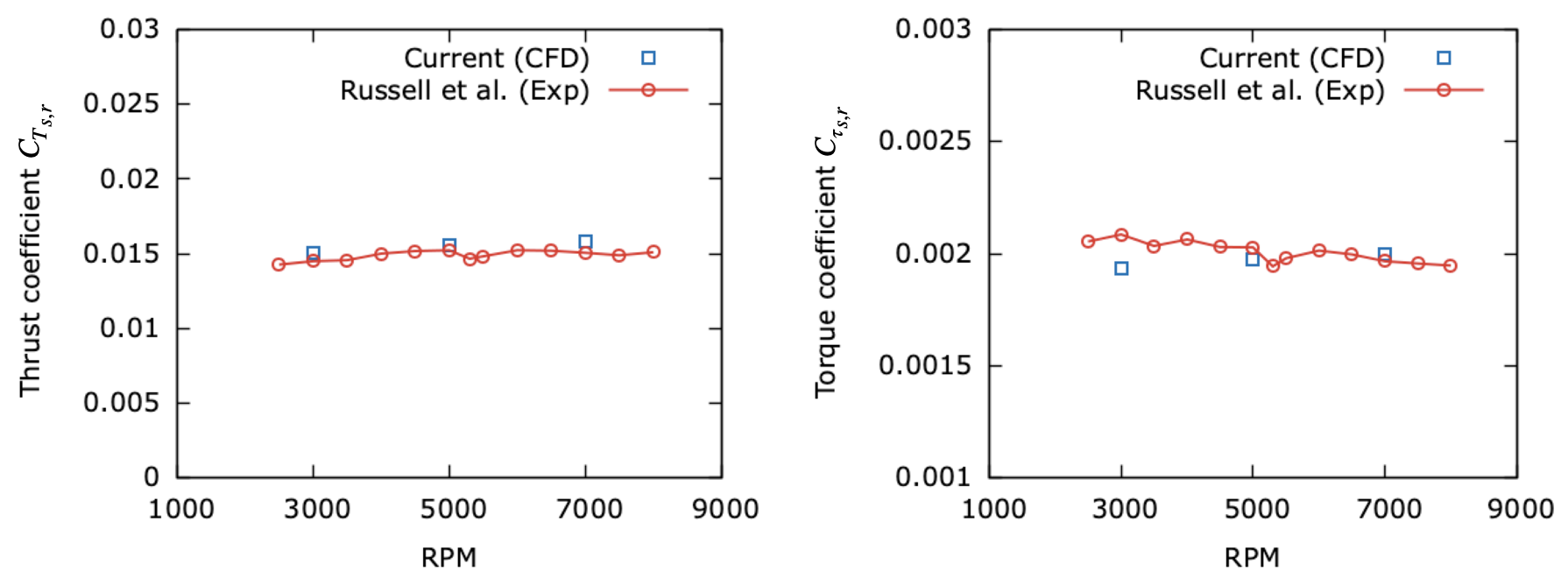}
	\caption{The thrust $C_{T_{s,r}}$ and torque $C_{\tau_{s,r}}$ coefficients for the isolated single rotor in comparison to previous experimental data \cite{Russell2018}.}
	\label{fig:experiments_single}
\end{figure}

The rotor thrust $T_{s,r}$ and torque $\tau_{s,r}$ are compared with previous experimental data from Russell et al. \cite{Russell2018} in Fig. \ref{fig:experiments_single}. 
The rotor thrust and torque coefficients are defined as $C_{T_{s,r}} = T_{s,r}/\rho A (\Omega R)^2$ and $C_{\tau_{s,r}} = \tau_{s,r}/ \rho A (\Omega R)^2 R$, where $A$ is the rotor disk area. 
As shown in Fig. \ref{fig:experiments_single}, the predicted $C_{T_{s,r}}$ and $C_{\tau_{s,r}}$ in the computations are matched well with the test data of Russell et al. \cite{Russell2018} at three rotor speeds. 

\begin{figure}[htb!] \centering
	\includegraphics[width=0.5\textwidth]{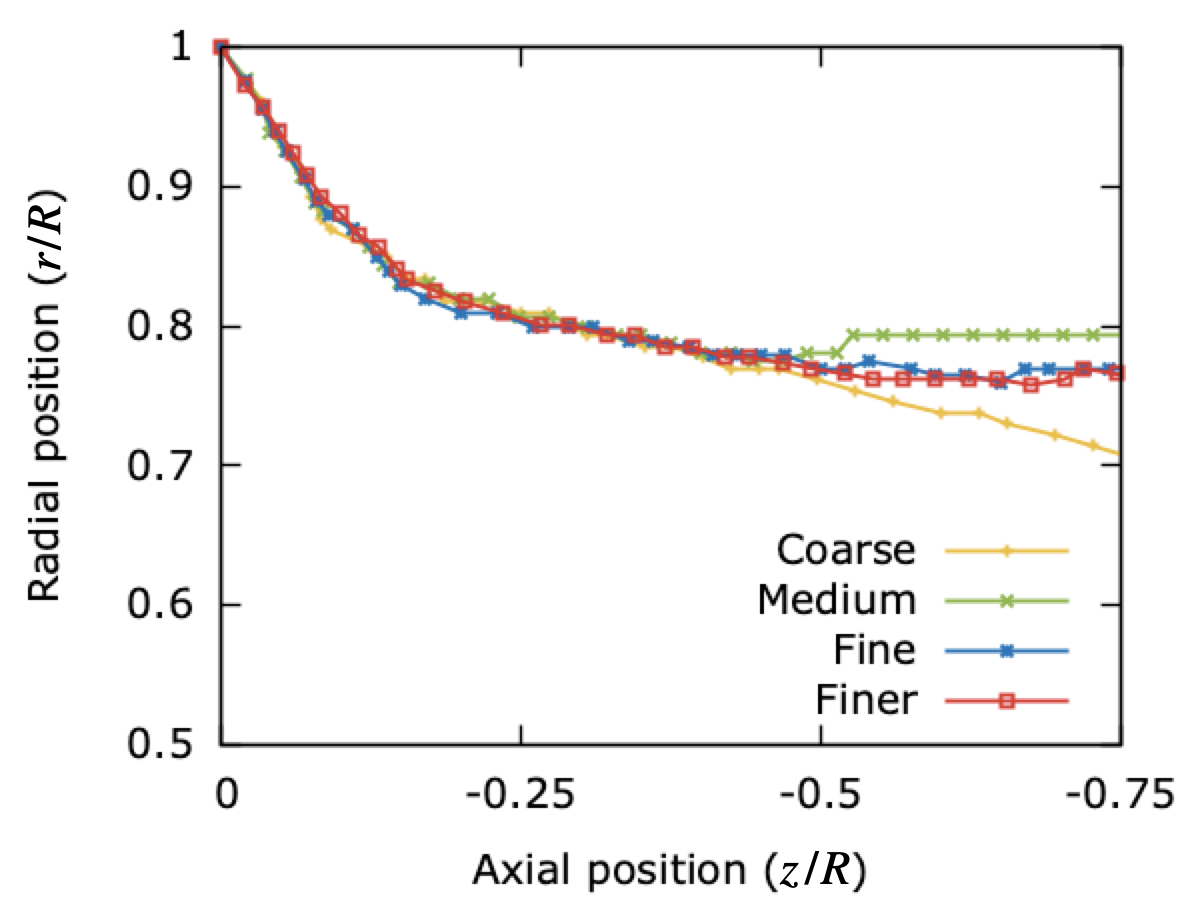}
	\caption{The position of the tip vortices for the isolated single rotor with four grid resolutions at the rotor speed $\Omega$ = 5000 RPM.}
	\label{fig:convergence_single_wake}
\end{figure}

The comparison of rotor wakes in four grid resolutions is shown in Fig. \ref{fig:convergence_single_wake}. 
The positions of the tip vortices along the radial and axial direction ($r, z$) are phase-averaged over two rotor revolutions. 
As the grid resolution is refined, the rotor wake tends to converge towards that of the finest grid.
The position of tip vortices within the grids Fine and Finer is nearly identical.
It can be assumed that the grid Fine adequately resolves the flow fields with negligible discretization error.
Therefore, computational results from the grid Fine are reported in this paper, unless otherwise stated.

\begin{figure}[htb!] \centering
	\includegraphics[width=\textwidth]{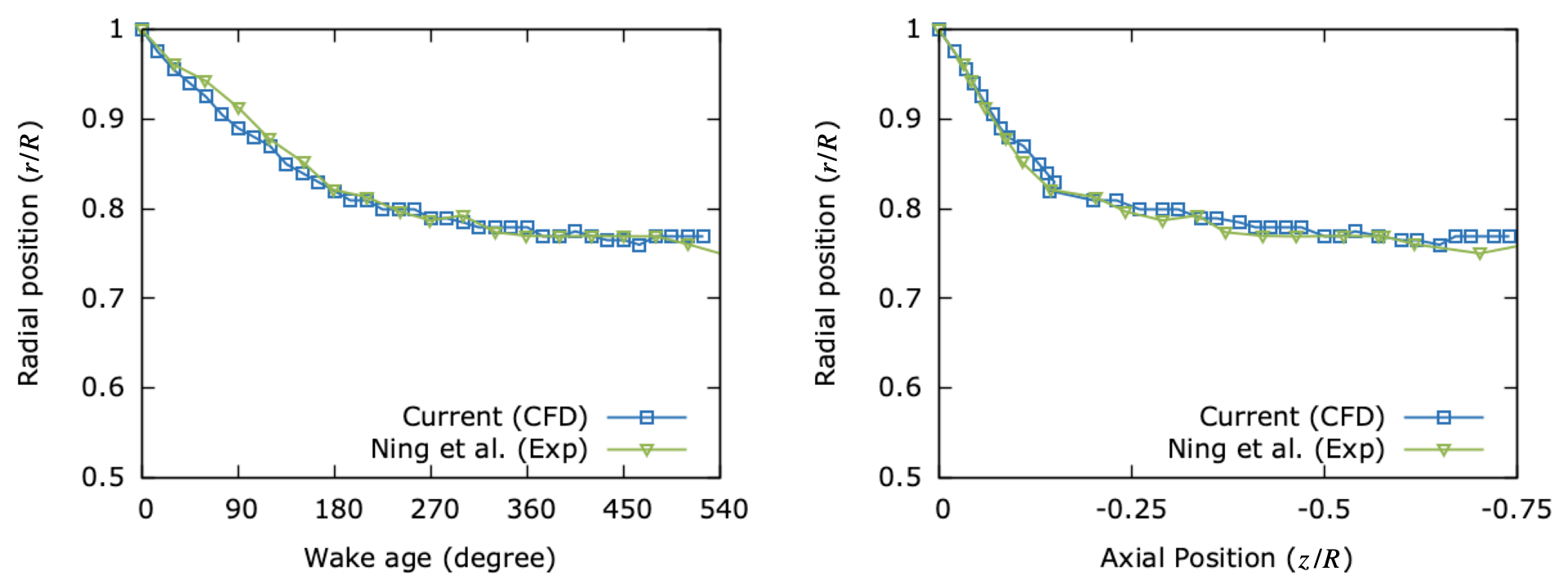}
	\caption{The positions of the tip vortices for the isolated single rotor in comparison to previous experimental data \cite{Ning2018}.}
	\label{fig:experiments_single_wake}
\end{figure}

The positions of the tip vortices in the isolated single rotor is compared with previous experimental data \cite{Ning2018}, as shown in Fig. \ref{fig:experiments_single_wake}. 
Tip vortex positions from the current computations are phase-averaged over two rotor revolutions.
The positions of tip vortices in current computations align closely with experimental measurement.

\begin{figure}[htb!] \centering
	\includegraphics[width=0.95\textwidth]{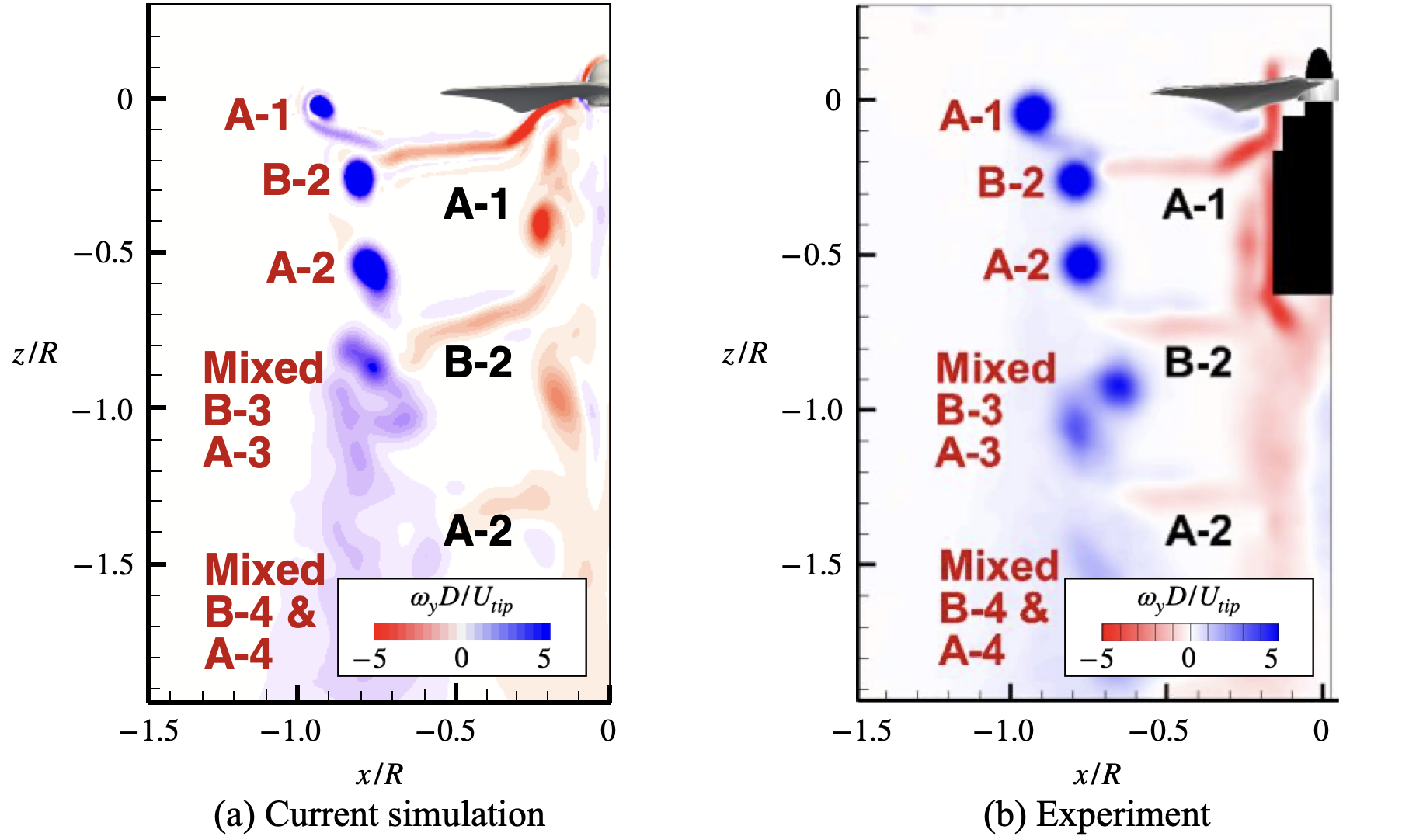}
	\caption{Phase-averaged vorticity field at the rotor azimuthal angle $\Psi = 60^\circ$ in comparison to previous experimental data \cite{Ning2018}.}
	\label{fig:experiments_single_vorticity}
\end{figure}

Phase-averaged vorticity fields at the rotor azimuthal angle $\Psi$ = 60$^\circ$ are also compared with experimental data \cite{Ning2018} as shown in Fig. \ref{fig:experiments_single_vorticity}.
Label A marks tip vortices induced from the blade shown in Fig. \ref{fig:experiments_single_vorticity}, and label B denotes tip vortices from the other blade.
Label 1 indicates the current blade cycle, and labels 2 or higher indicate previous cycle numbers.
Labels for tip vortices and vortex sheets are colored with red and black, respectively.
Three rotor tip vortices (labelled as A-1, B-2, and A-2 with red color in Fig. \ref{fig:experiments_single_vorticity}) are positioned in similar locations observed in the experiment.
Mixed tip vortices (labelled as Mixed B-3 and A-3) are also qualitatively captured in the current simulation.
Therefore, the current numerical approach using the overset mesh reproduces relevant experimental data. 
The validation of the current approach for the quadcopter case is discussed in the following section.

\subsection{Quadcopter \label{sec:quadcopter}}

The quadcopter configuration is simulated with the grid Fine, which is validated with the isolated single rotor in Section \ref{sec:single}.
The near-body rotor grid in the current configuration is the same as in the isolated single rotor case, and the off-body grid has structured isotropic cells uniformly allocated around the near-body rotor grid, which is similarly used in the single rotor case.

\begin{figure}[htb!] \centering
	\includegraphics[width=0.55\textwidth]{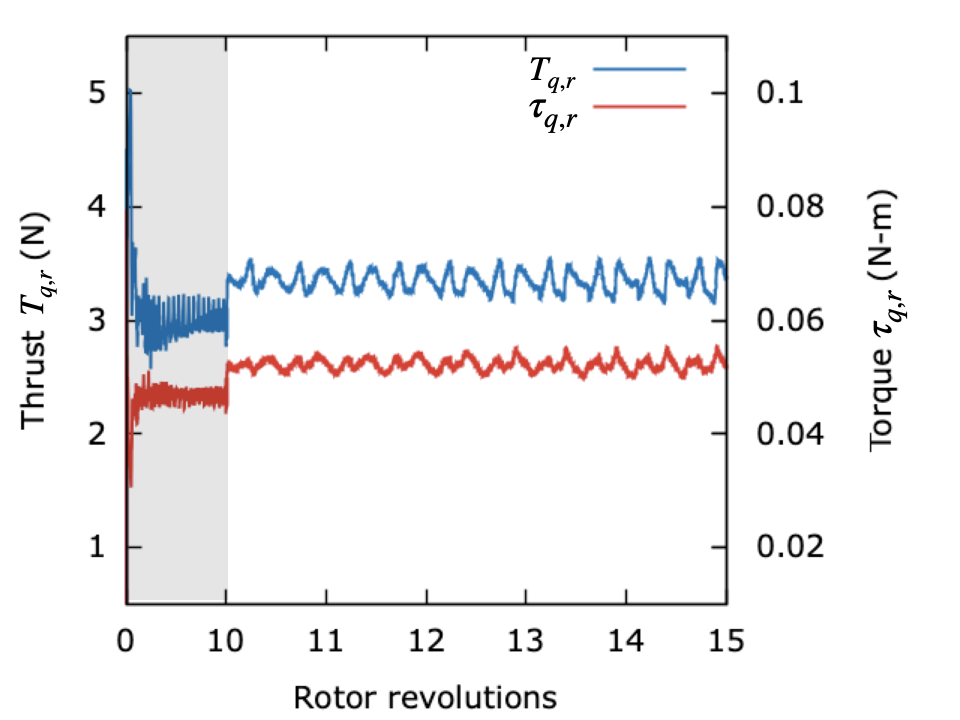}
	\caption{Revolution convergence of the quadcopter rotor thrust $T_{q,r}$ and torque $\tau_{q,r}$ at the rotor speed $\Omega$ = 5000 RPM.}
	\label{fig:convergence_complete}
\end{figure}

In the current unsteady computations, a quasi-steady state of the thrust $T_{q,r}$ and torque $\tau_{q,r} $ for the quadcopter rotor (denoted by $(q,r)$) is achieved, as shown in Fig. \ref{fig:convergence_complete}. 
Ten rotor revolutions are initially simulated with a relatively large time step equivalent to $\Delta \theta = 2.5\degree$ for rapid flow development (the grey region in Fig. \ref{fig:convergence_complete}). 
The time step is reduced to $\Delta \theta = 0.125\degree$, as in the case of the isolated single rotor. 
The quadcopter rotor $T_{q,r}$ and $\tau_{q,r}$ converge after 11 rotor revolutions. 
The quadcopter rotors experience fluctuating $T_{q,r}$ and $\tau_{q,r}$ with a major periodicity of half a revolution, unlike the isolated single rotor. 
The differences in the time-averaged thrust and torque between the final two rotor revolutions are only 0.05\%.
Therefore, the time-averaged thrust and torque are obtained from the 13th and 14th rotor revolutions. 

\begin{figure}[htb!] \centering
	\includegraphics[width=\textwidth]{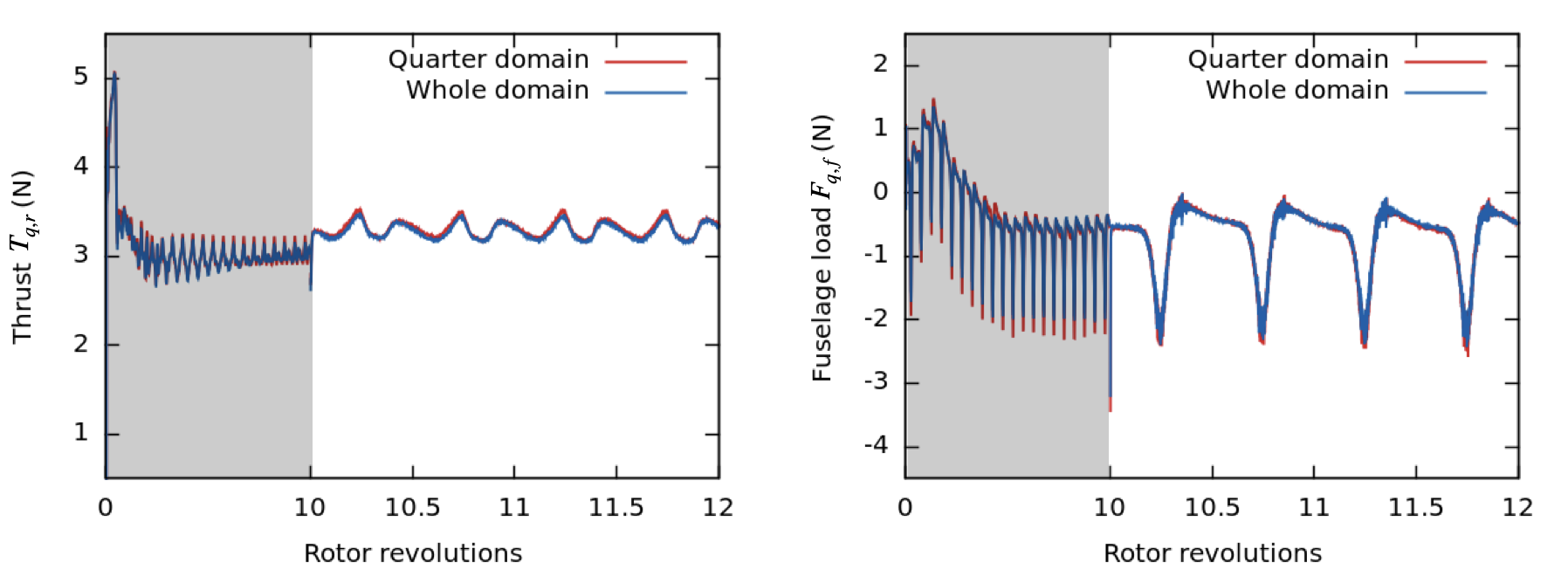}
	\caption{Revolution convergence of the quadcopter rotor thrust $T_{q,r}$ and the fuselage load $F_{q,f}$ for the two computational domains at the rotor speed $\Omega$ = 5000 RPM.} 
	\label{fig:convergence_complete_vs_quarter}
\end{figure}

The comparison of the quadcopter rotor thrust $T_{q,r}$ and fuselage load $F_{q,f}$ is presented for  two computational domains, i.e., the whole domain and the quarter domain, as shown in Fig. \ref{fig:convergence_complete_vs_quarter}.
Both the domains provide nearly identical $T_{q,r}$ and $F_{q,f}$, capturing unsteady airloads.
The differences in the time-averaged $T_{q,r}$ and $F_{q,f}$ between the two domains amount to only 0.8\%.
Thus, it can be thought that the quarter domain, with its symmetric boundaries, effectively resolves the unsteady aerodynamics. 
Consequently, computational results with the quarter domain are reported in this paper for the quadcopter simulation, unless otherwise mentioned.

\begin{figure}[htb!] \centering
	\includegraphics[width=.5\textwidth]{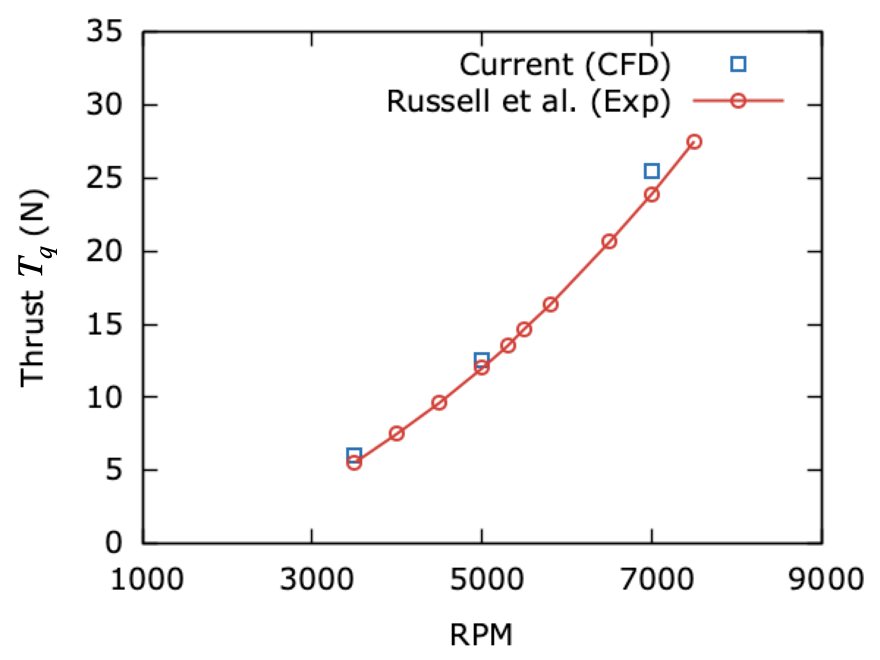}
	\caption{The thrust $T_q$ for the quadcopter in comparison to previous experimental data \cite{Russell2018}.}
	\label{fig:experiment_complete}
\end{figure}

The quadcopter thrust $T_q$ is compared with experimental studies by Russell et al. \cite{Russell2018} as shown in Fig. \ref{fig:experiment_complete}.
The quadcopter thrust $T_q$ is calculated as the sum of four quadcopter rotor thrusts and the quadcopter fuselage load $T_q = 4T_{q,r} + F_{q,f}$. The predicted $T_q$ agrees well with previous experimental data from Russell et al. \cite{Russell2018}. 

\begin{figure}[htb!] \centering
	\includegraphics[width=.8\textwidth]{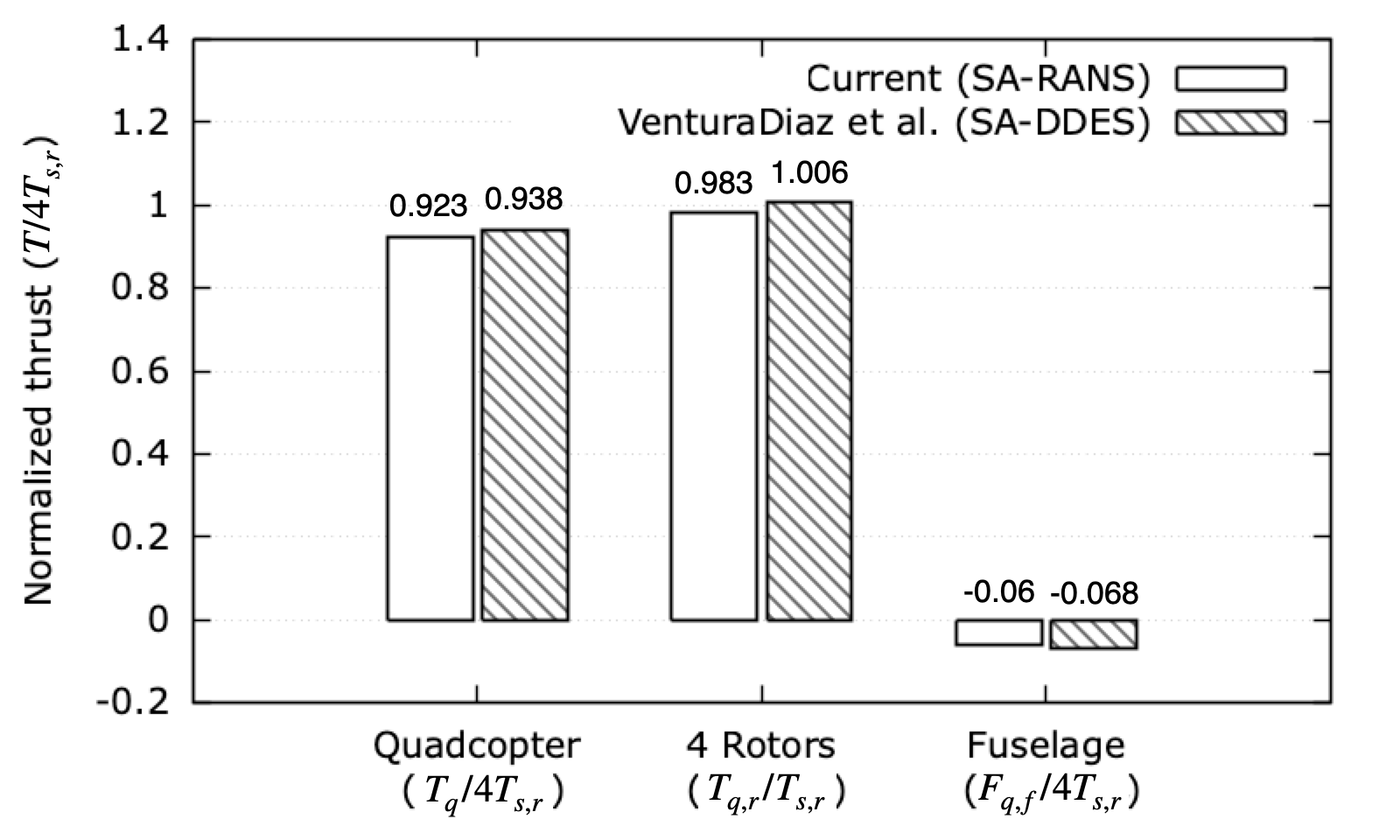}
	\caption{Normalized thrust for the quadcopter components in comparison to  previous computational data \cite{VenturaDiaz2018}.}
	\label{fig:barComponent_thrust_previous}
\end{figure}

The normalized thrust of the quadcopter components is compared with previous higher-fidelity computations \cite{VenturaDiaz2018}, as shown in Fig. \ref{fig:barComponent_thrust_previous}. 
The single-rotor counterpart is used for the normalization. The overall quadcopter thrust is reduced about 7.7\% compared to the isolated single rotor thrust $4T_{s,r}$. 
Most of the thrust reduction is related with the quadcopter fuselage load, which accounts for 6\% of the isolated single rotor thrust. 
This reduction in the quadcopter thrust is also reported in another computational study \cite{VenturaDiaz2018}, which used the same quadcopter with the delayed-detached-eddy simulation (DDES).  
Thus, the current numerical approach can predict well the airloads in the range of the literature data.

The validation of the current computational method for the quadcopter configuration is confirmed through the comparison of experimental and computational data in literature. 
The quadcopter simulation demonstrates unsteady airloads significantly different to the isolated single rotor, which will be further investigated in the following section.

\section{Results\label{sec:unsteady}}

\begin{figure}[htb!] \centering
	\includegraphics[width=\textwidth]{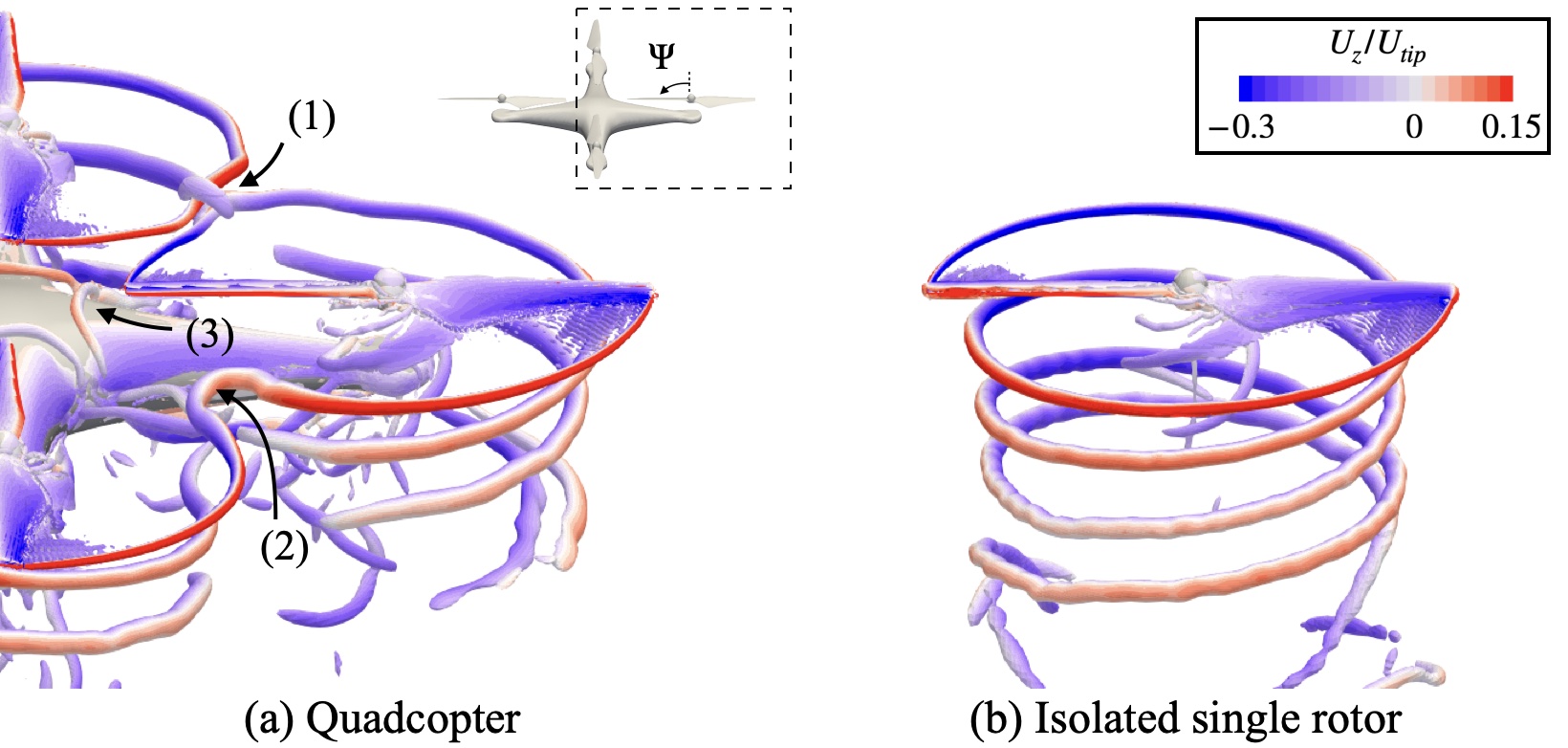}
	\caption{Iso-surfaces of the $Q$-criterion ($Q = 1\times10^6~\rm{s^{-2}}$) for the isolated single rotor and the quadcopter ($\Psi = 90\degree$) at the rotor speed $\Omega$ = 5000 RPM.}
	\label{fig:Qcrit_quadcopter_vs_single}
\end{figure}

Complex vortical structures emerge around the quadcopter due to the interactional effects between the adjacent rotors and the fuselage.
Tip vortices around the quadcopter evolve into distinct vortical structures, denoted as (1), (2), and (3) in Fig. \ref{fig:Qcrit_quadcopter_vs_single}.
Aerodynamic interaction between adjacent rotors and the fuselage can influence the tip vortices, resulting in distinct vortical structures in contrast to the helical vortices in the isolated single rotor.
At the rotor azimuthal angles $\Psi = 45\degree$ and $135\degree$, where the two rotors are closely located, tip vortices from adjacent rotors merge into $\Omega$-shaped vortical structures (see (1) and (2) in Fig. \ref{fig:Qcrit_quadcopter_vs_single}). 
These $\Omega$-shaped vortical structures have also been reported in previous side-by-side rotor simulation \cite{Sagaga2021} as a result of the interactional effect between two rotors.
The blockage of the fuselage around $\Psi = 90\degree$ disrupts the formation of tip vortices, causing a stretched vortex tube over its fuselage arm (see (3) in Fig. \ref{fig:Qcrit_quadcopter_vs_single}).

\begin{figure}[htb!] \centering
	\includegraphics[width=\textwidth]{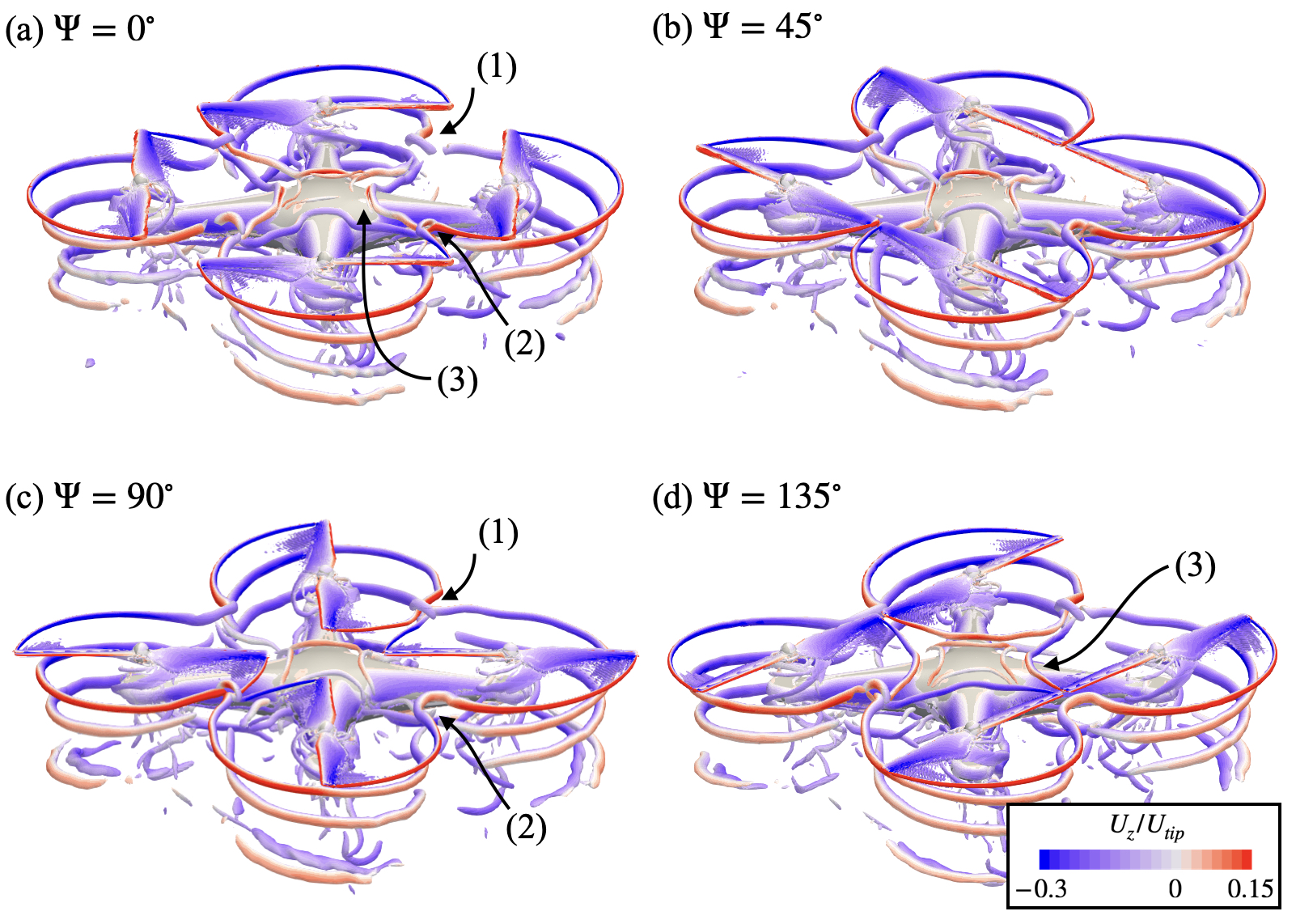}
	\caption{Evolution of vortical structures visualized with iso-surfaces of the $Q$-criterion ($Q = 1\times 10^6~\rm{s^{-2}}$) for the quadcopter with the variation of the azimuthal angle $\Psi$ defined for the right rotor.}
	\label{fig:Qcrit_quadcopter_angles}
\end{figure}

Vortical structures are evolved in time as shown in Fig. \ref{fig:Qcrit_quadcopter_angles} where four azimuthal angles $\Psi$ are selected for the flow visualization.
These angles correspond to the initial state ($\Psi = 0\degree$), rotor-rotor interaction states ($\Psi = 45\degree$ and $135\degree$), and the rotor-fuselage interaction state ($\Psi = 90\degree$).
After $\Psi = 45\degree$, where the rotor-rotor interaction occurs, tip vortices from adjacent rotors are merged into the $\Omega$-shaped vortex (1).
Similarly, the vortex (2) is formed between two rotors after $\Psi = 135\degree$ as shown in Fig. \ref{fig:Qcrit_quadcopter_angles} (d). 
The stretched vortex tube (3) is caused by the obstruction of the fuselage on the helical tip vortex. 
Such vortices are accumulated above the fuselage as the rotors revolve. 

\begin{figure}[htb!] \centering
	\includegraphics[width=\textwidth]{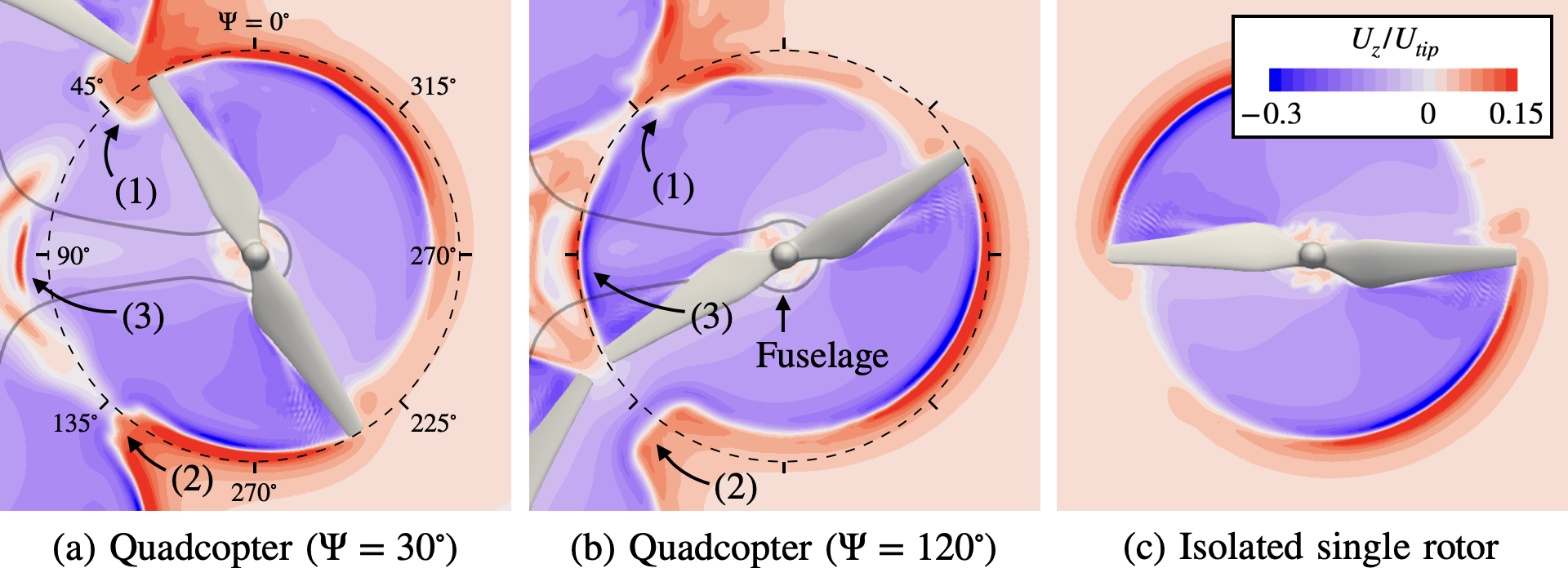}
	\caption{Instantaneous axial velocity fields ($U_z/U_{tip}$) on the plane $z=-0.5 c_{tip}$ at the rotor speed $\Omega$ = 5000 RPM.}
	\label{fig:rotor-rotor_interaction_UZ}
\end{figure}

Flow fields around the quadcopter are influenced by the evolution of major vortical structures.
The instantaneous axial velocity $U_z$ is compared between the quadcopter and the isolated single rotor on the rotor disk plane at $z=-0.5 c_{tip}$ as shown in Fig. \ref{fig:rotor-rotor_interaction_UZ}.
The $\Omega$-shaped vortex (1) distorts the boundary of the rotor wake around $\Psi = 45\degree$.
This distorted wake induces the upwash and downwash flows in the outboard section of the rotor blade before and after $\Psi = 45\degree$, respectively.
Similarly, the $\Omega$-shaped vortex (2) induces the downwash and upwash flows before and after $\Psi = 135\degree$.
In addition, the fuselage also disrupts the convection of the rotor-induced flow, which stretches tip vortices (3) around $\Psi = 90\degree$. 

\begin{figure}[htb!] \centering
	\includegraphics[width=\textwidth]{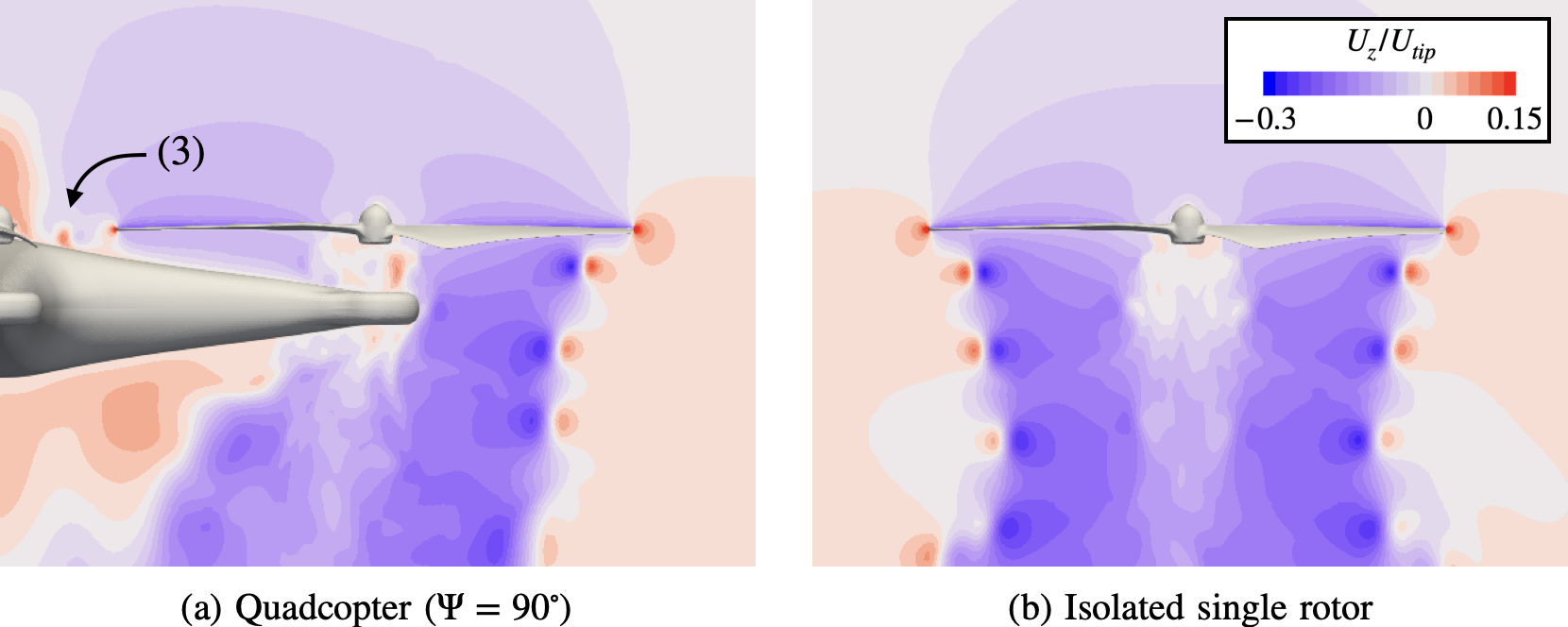}
	\caption{Instantaneous axial velocity fields ($U_z/U_{tip}$) on the plane $y=0$ at the rotor speed $\Omega$ = 5000 RPM.}
	\label{fig:rotor-fuselage_interaction_UZ}
\end{figure}

The rotor-fuselage interaction has a significantly impact on the instantaneous axial velocity $U_z$ on the plane $y = 0$. 
The rotor blade is positioned just above the fuselage ($\Psi = 90\degree$) as shown in Fig. \ref{fig:rotor-fuselage_interaction_UZ}.
The fuselage obstructs a portion of the rotor wake, in contrast to the isolated single rotor.
Therefore, the tip vortex (3) remains above the fuselage, instead of being convected away as in the isolated single rotor case. 
These accumulated tip vortices generate the upwash flow as the rotor revolves.
In addition, the obstruction of the wake causes recirculation below the fuselage, which bends the rotor wake towards the center of the quadcopter.

\begin{figure}[htb!] \centering
	\includegraphics[width=\textwidth]{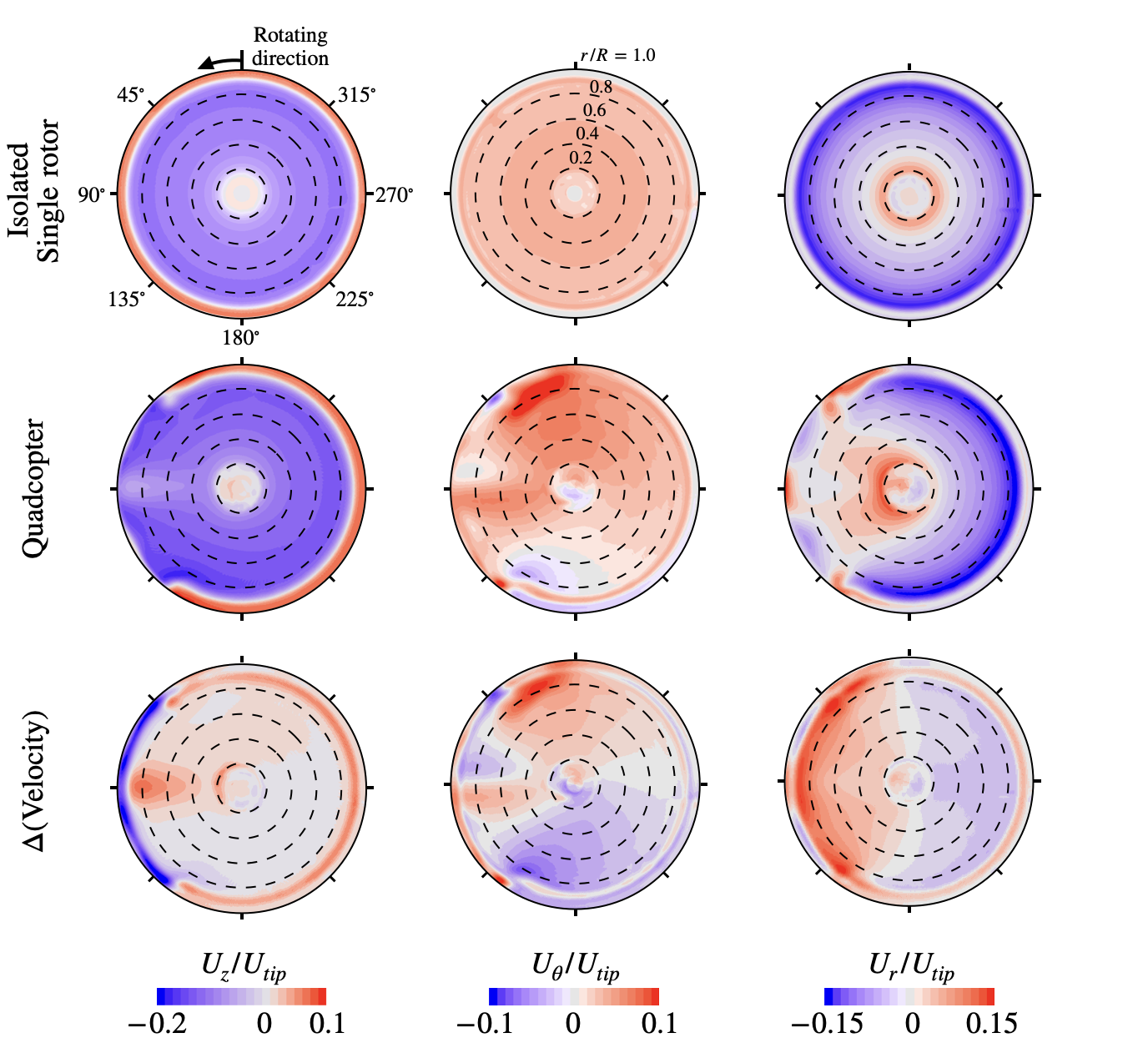}
	\caption{Time-averaged velocity fields on the plane $z=-0.5 c_{tip}$ for the isolated single rotor (top), the quadcopter rotor (center), and the variations between two rotors (bottom) at the rotor speed $\Omega$ = 5000 RPM.}
	\label{fig:time_averaged_meanflow}
\end{figure}

The time-averaged induced velocities $U_z$, $U_\theta$, and $U_r$ are compared between the quadcopter and the isolated single rotor on the rotor disk plane at $z=-0.5 c_{tip}$, as shown in Fig. \ref{fig:time_averaged_meanflow}. 
To emphasize interactional effects between two configurations, the difference of the induced velocity is calculated as $\Delta (U/U_{tip}) = (U/U_{tip})_{q,r} - (U/U_{tip})_{s,r}$.
In the quadcopter, the induced flow is significantly modulated due to aforementioned major vortical structures related to rotor-rotor interaction ($\Psi =  45\degree$ and $135\degree$) and rotor-fuselage interaction ($\Psi = 90\degree$).

In the quadcopter, the wake boundary is distorted by $\Omega$-shaped vortices around the rotor angles $\Psi = 45\degree$ and $135\degree$.
Therefore, the upwash flow in the outboard section ($r/R \approx 0.9$) vanishes in the range $45\degree \lesssim \Psi \lesssim 135\degree$, which results in the decreased axial flow $U_z$.
Furthermore, $\Omega$-shaped vortices induce positive azimuthal $U_\theta$ flows around $\Psi = 45\degree$, and negative $U_\theta$ around $\Psi = 135\degree$.

Around $\Psi = 90\degree$, the fuselage obstructs the rotor wake, varying $U_z$ and $U_\theta$. 
Since the presence of the fuselage bends the rotor wakes towards the center of the vehicle, $\Delta U_\theta$ is overall positive in the range $270\degree \lesssim \Psi \lesssim 90\degree + 360\degree$ and negative in the range $90\degree \lesssim \Psi \lesssim 270\degree$.
The inclined rotor wake also induces positive $\Delta U_r$ around the first and second quadrant ($0\degree \lesssim \Psi \lesssim 180\degree$), and negative $\Delta U_r$ around the third and fourth quadrant ($180\degree \lesssim \Psi \lesssim 360\degree$) of the rotor disk.

\begin{figure}[htb!] \centering
	\includegraphics[width=\textwidth]{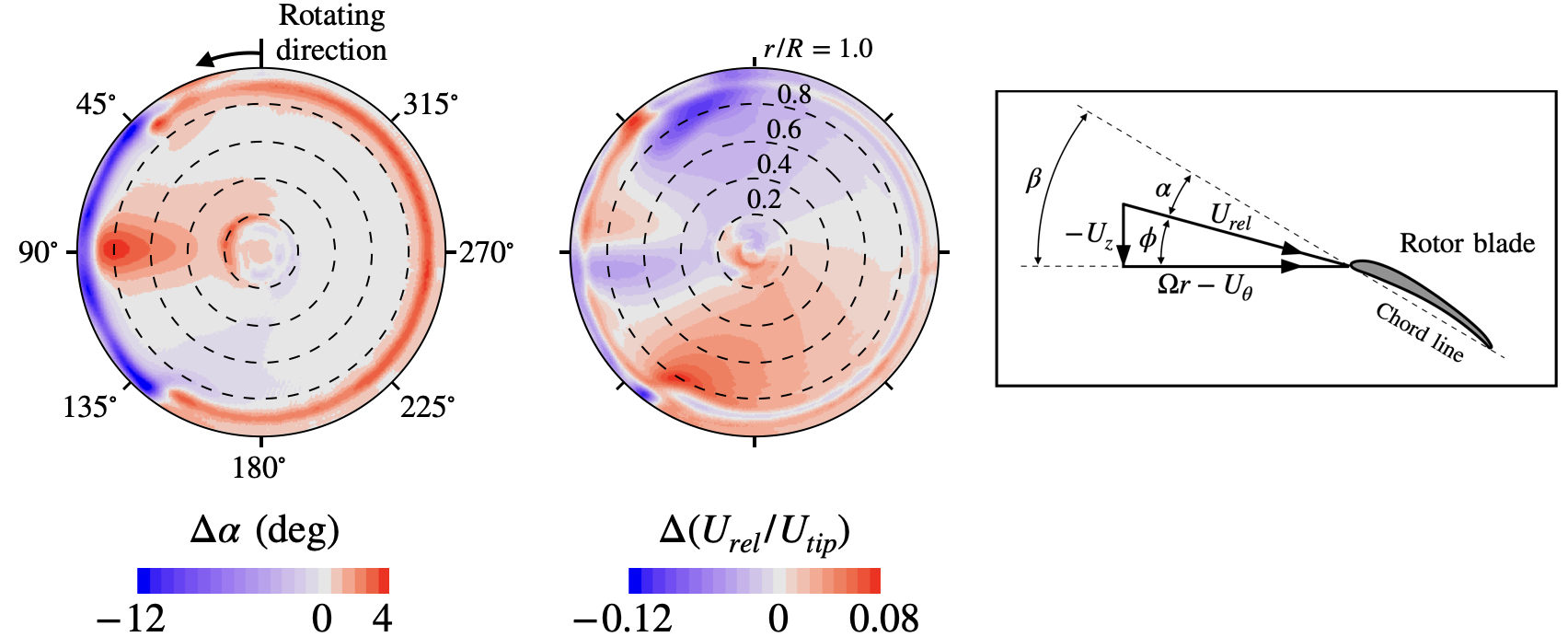}
	\caption{Difference of the time-averaged effective angle of attack $\alpha$ and the relative inflow velocity $U_{rel}$ between the quadcopter and isolated single rotor at the rotor speed $\Omega$ = 5000 RPM.}
	\label{fig:time_averaged_inflow_deviation}
\end{figure}

The rotor-induced flow changes both the effective angle of attack $\alpha$ and the relative inflow velocity $U_{rel}$ for a rotor blade as shown in Fig. \ref{fig:time_averaged_inflow_deviation}.
The relative parameters $\alpha$ and $U_{rel}$ are calculated as $\alpha = \beta - \phi$ and $U_{rel}= (\Omega r - U_{\theta})\hat{e}_\theta + (-U_z)\hat{e}_z$, where $\phi$ is the relative inflow angle.

Positive $\Delta U_z$ in Fig. \ref{fig:time_averaged_meanflow} reduces the relative axial inflow velocity $(-U_z)\hat{e}_z$ and the relative inflow angle $\phi$, which increases $\alpha$ for the rotor blade in Fig. \ref{fig:time_averaged_inflow_deviation}. 
In contrast, negative $\Delta U_z$ decreases $\alpha$ in comparison with the isolated single rotor.
Therefore, the distribution of the difference in the effective angle of attack $\Delta \alpha$ in Fig. \ref{fig:time_averaged_inflow_deviation} is similar to that of $\Delta U_z$ in Fig. \ref{fig:time_averaged_meanflow}.

Similarly, positive $\Delta U_\theta$ in Fig. \ref{fig:time_averaged_meanflow} reduces the relative incoming velocity $(\Omega r - U_{\theta})\hat{e}_\theta$, eventually decreasing the inflow velocity $U_{rel}$.
On the contrary, negative $\Delta U_\theta$ increases the $(\Omega r - U_{\theta})\hat{e}_\theta$ and $U_{rel}$.
As a result, the distribution of the difference in the inflow velocity $\Delta U_{rel}$ in Fig. \ref{fig:time_averaged_inflow_deviation} is inversely similar to that of $\Delta U_\theta$ in Fig. \ref{fig:time_averaged_meanflow}.
Therefore, the rotor blade encounters higher $U_{rel}$ mostly in the second and third quadrants, and lower $U_{rel}$ mostly in the fourth and first quadrants.

\begin{figure}[htb!] \centering
	\includegraphics[width=0.90\textwidth]{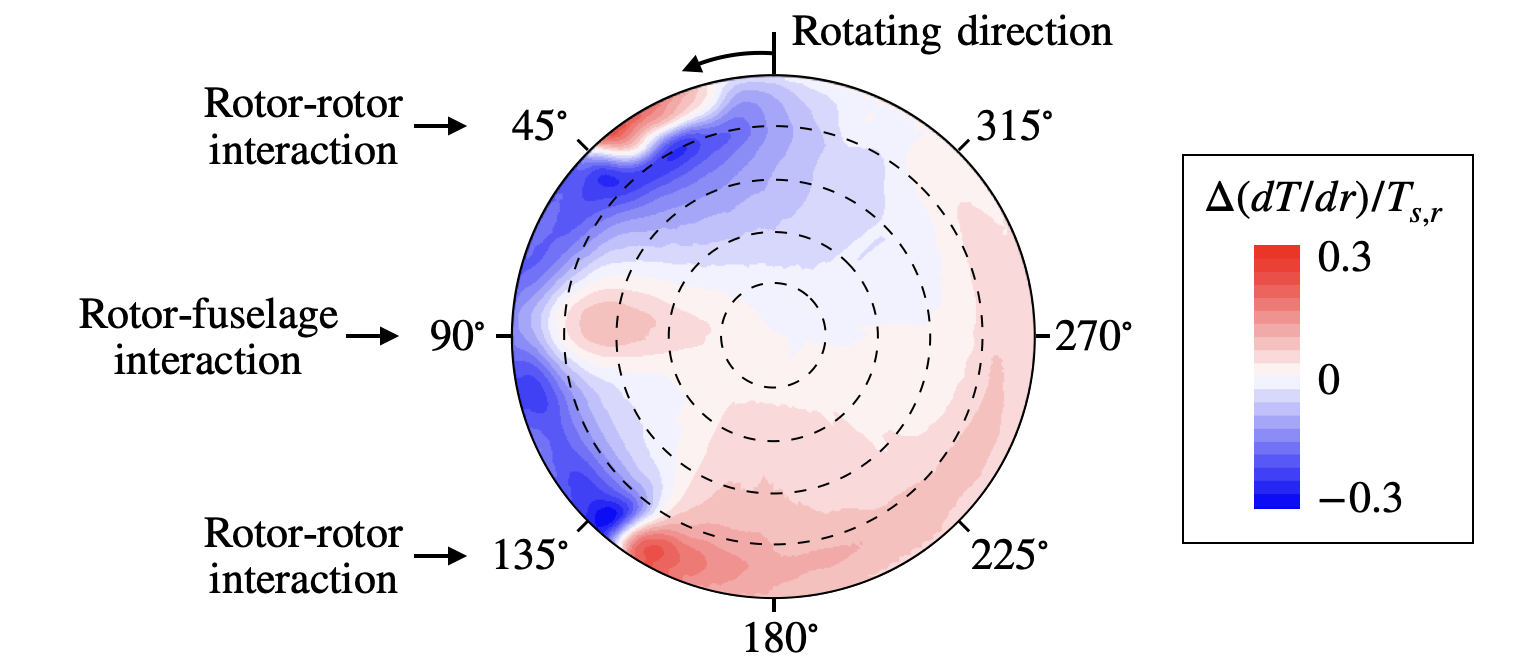}
	\caption{Difference of the sectional thrust between the quadcopter rotor and isolated single rotor $\Delta(dT/dr) = (dT/dr)_{q,r} - (dT/dr)_{s,r}$ over the one revolution at the rotor speed $\Omega$ = 5000 RPM.}
	\label{fig:sectionalThrust_5000RPM_difference}
\end{figure}

Variations in the effective angle of attack $\alpha$ and the relative inflow velocity $U_{rel}$ in Fig. \ref{fig:time_averaged_inflow_deviation} change a sectional thrust $dT/dr$ on the quadcopter rotor blade as shown in Fig. \ref{fig:sectionalThrust_5000RPM_difference}.
The difference of $dT/dr$ between the quadcopter rotor and the isolated single rotor is calculated as $\Delta (dT/dr)$ = $(dT/dr)_{q,r} - (dT/dr)_{s,r}$.

Around the rotor angles $\Psi = 45\degree$ and $135\degree$ (denoted as the rotor-rotor interaction in Fig. \ref{fig:sectionalThrust_5000RPM_difference}), $dT/dr$ varies in the outboard section $r/R > 0.8$, due to the interaction with $\Omega$-shaped vortical structures.
Before reaching $\Psi = 45\degree$, the rotor blade experiences locally increased $dT/dr$ in the outboard section, due to the increased $\alpha$ induced by the upwash flow from the $\Omega$-shaped vortex.
After $\Psi = 135\degree$, $dT/dr$ increases in the outboard section due to the increased $\alpha$.
Within the range $45\degree \lesssim \Psi \lesssim 135\degree$, the rotor blade encounters a reduced $dT/dr$ in the outboard section, resulting from the decreased $\alpha$. 

In the inboard section of the quadcopter rotor $r/R < 0.8$, $dT/dr$ increases in the second and third quadrants ($90\degree \lesssim \Psi \lesssim 270\degree$) primarily because of a large relative velocity (see the corresponding region of $\Delta U_{rel}>0$ in Fig. \ref{fig:time_averaged_inflow_deviation}). On the other hand, $dT/dr$ decreases in the fourth and first quadrants ($270\degree \lesssim \Psi \lesssim 90\degree + 360\degree$) primarily because of a low relative velocity $U_{rel}$ there.
Additionally, the rotor blade experiences higher $dT/dr$ in the inboard section around $\Psi = 90\degree$ (denoted as the rotor-fuselage interaction in Fig. \ref{fig:sectionalThrust_5000RPM_difference}), due to the increased $\alpha$ (see $\Delta \alpha > 0$ around $\Psi = 90\degree$ in Fig. \ref{fig:time_averaged_inflow_deviation}) resulting from the presence of the fuselage.

\begin{figure}[htb!] \centering
	\includegraphics[width=.6\textwidth]{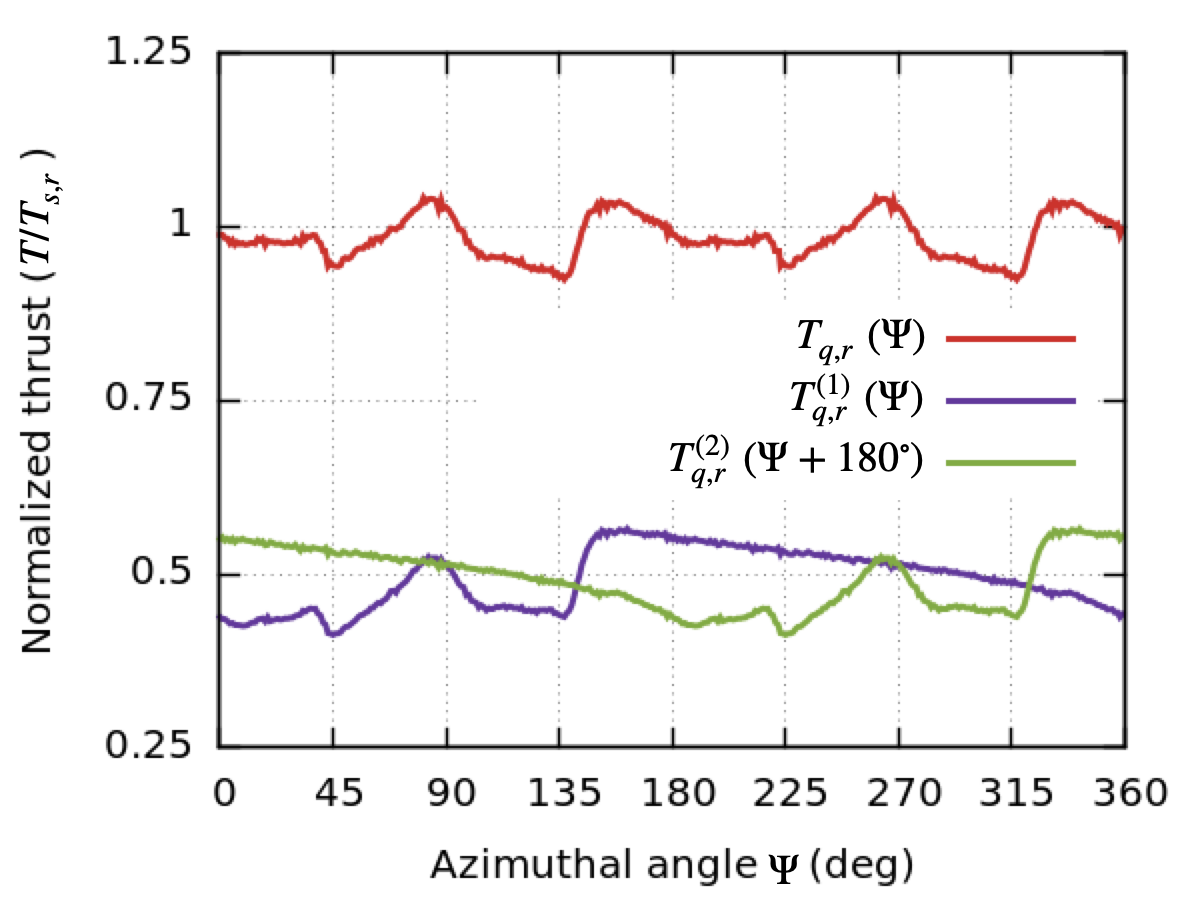}
	\caption{Vertical forces on the quadcopter rotor $T_{q,r}$ and the two rotor blades ($T^{(1)}_{q,r}$ and $T^{(2)}_{q,r}$) over the one rotor revolution at the rotor speed $\Omega =$ 5000 RPM.}
	\label{fig:blade_thrust_1rev}
\end{figure}

The integrated thrust $T$ over the quadcopter rotor fluctuates significantly as shown in Fig. \ref{fig:blade_thrust_1rev}.
Thrust of each blade is denoted as $T^{(1)}_{q,r}$ and $T^{(2)}_{q,r}$ for the two-blade rotor. Note that the thrust on the second blade is shifted by the phase $180\degree$ because the azimuthal angle $\Psi$ is defined for the first blade. 
The quadcopter rotor thrust $T_{q,r}$ is the sum of the thrust forces from the two rotor blades, $T_{q,r} = T^{(1)}_{q,r} + T^{(2)}_{q,r}(\Psi + 180\degree)$.

The blade thrust $T^{(1)}_{q,r}$ varies significantly in the first half of the revolution due to the rotor-rotor and rotor-fuselage interaction. In $0 \leq \Psi \lesssim 45 \degree$, $T^{(1)}_{q,r}$ does not change much because the high $dT/dr$ in the outboard section is mostly balanced with the low $dT/dr$ in the inboard section. In $45\degree \lesssim \Psi \lesssim 90 \degree$, $T^{(1)}_{q,r}$ increases due to the increasing $dT/dr$ in the inboard section mostly caused by the fuselage blockage effect. The blockage effect vanishes after $\Psi \simeq 90 \degree$, so $T^{(1)}_{q,r}$ drops. The thrust $T^{(1)}_{q,r}$ increases around $\Psi \simeq 135 \degree$ because of the large $dT/dr$ on the overall blade (see Fig. \ref{fig:sectionalThrust_5000RPM_difference}), which is related to the $\Omega$-shaped vortex phenomena (see Fig. \ref{fig:rotor-rotor_interaction_UZ}) formed by the rotor-rotor interaction. After $\Psi \simeq 150\degree$, $T^{(1)}_{q,r}$ gradually decreases until $\Psi \simeq  360\degree$, resulting from the diminished effect of the adjacent rotor and fuselage. The other blade experiences similar variations in $T^{(2)}_{q,r}$ with the phase difference of $180\degree$ to $T^{(1)}_{q,r}$. Consequently, the quadcopter thrust $T_{q,r}$ involves 4/rev thrust variations, resulting from the 2/rev thrust variation of each blade with the phase difference of $180\degree$.

\begin{figure}[htb!] \centering
	\includegraphics[width=.75\textwidth]{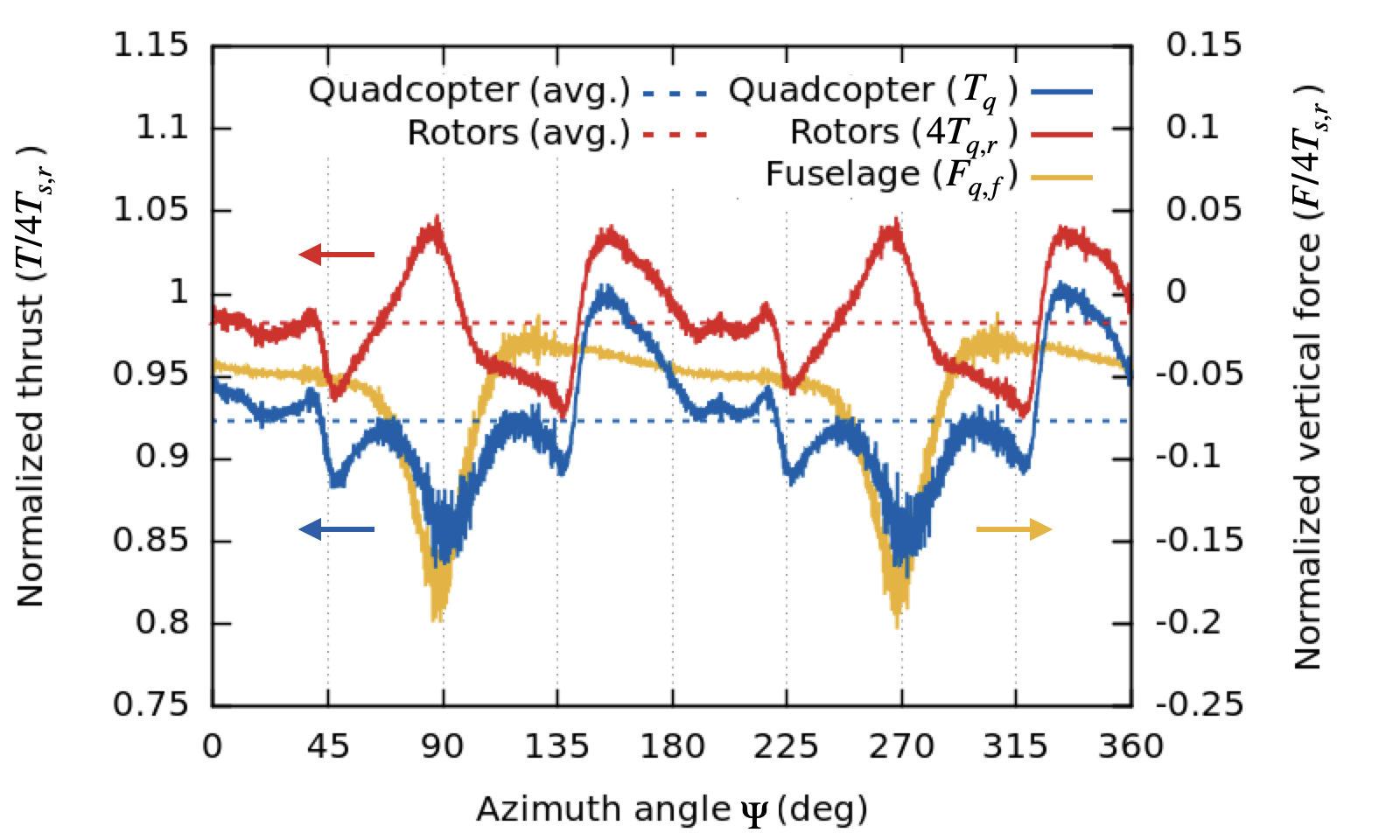}
	\caption{Vertical forces on the quadcopter $T_{q}$, the four rotors $4T_{q,r}$ and the fuselage $F_{q,f}$ over the one rotor revolution at the rotor speed $\Omega =$ 5000 RPM. Refer the right vertical axis for $F_{q,f}$. }
	\label{fig:component_thrust_1rev}
\end{figure}

The overall forces acting on the quadcopter and its components are shown in Fig. \ref{fig:component_thrust_1rev}.
The total thrust of the quadcopter is calculated as $T_q = 4T_{q,r} + F_{q,f}$ where $F_{q,f}$ is the vertical force generated by the fuselage.
The forces acting on the quadcopter components are normalized by the corresponding thrust of the single rotor $4T_{s,r}$ in Fig. \ref{fig:component_thrust_1rev}. 

The quadcopter experiences the minimum thrust $T_q$ around $\Psi = 90\degree$ and the maximum $T_q$ around $\Psi = 150\degree$ during half a rotor revolution (see blue curve in Fig. \ref{fig:component_thrust_1rev}). 
The minimum $T_q$ corresponds to the download of the quadcopter fuselage $F_{q,f}$, which is caused by the rotor wake (see the yellow curve in Fig. \ref{fig:component_thrust_1rev}).
The maximum $T_q$ is related to the peak of the rotor thrust $T_{q,r}$ (see the red curve in Fig. \ref{fig:component_thrust_1rev}) mainly caused by the rotor-rotor interaction, which is previously discussed with Fig. \ref{fig:blade_thrust_1rev}.
As a result, the major thrust variation occurs 2/rev. The valley-to-peak difference is about 15\% of the time-averaged thrust (the blue dashed line in Fig. \ref{fig:component_thrust_1rev}). Higher-frequency fluctuations of the total $T_q$ are primarily associated with the rotor thrust $T_{q,r}$, which will be discussed below with the frequency analysis.

\begin{figure}[htb!] \centering
	\includegraphics[width=\textwidth]{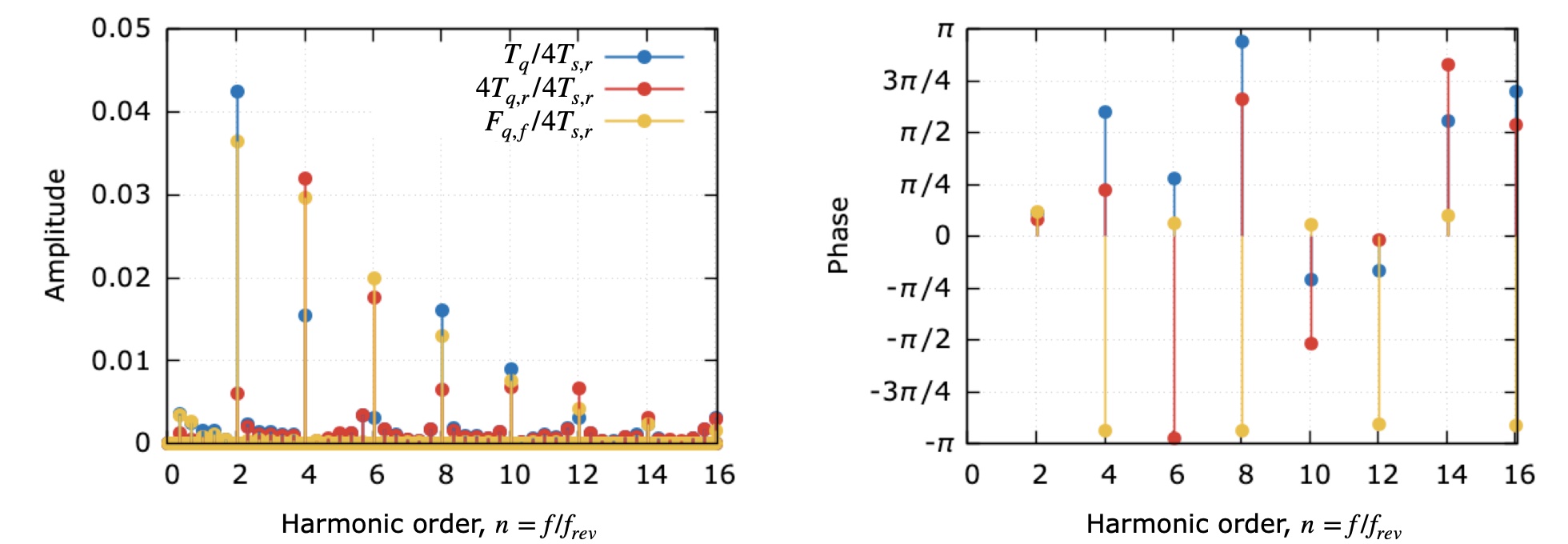}
	\caption{The frequency analysis of the quadcopter overall thrust $T_q$, the rotor thrust $4T_{q,r}$ and the fuselage force $F_{q,f}$.}
	\label{fig:component_thrust_FFT}
\end{figure}

Dominant frequencies of $T_q$, $4T_{q,r}$ and $F_{q,f}$ are identified using the Fourier analysis, as shown in Fig. \ref{fig:component_thrust_FFT}.
The harmonic order is obtained as $n = f/f_{rev}$, where $f$ is the frequency and $f_{rev}$ corresponds to the revolution rate $\Omega$ = 5000 RPM = 83.33 Hz.
Even harmonic orders are dominant because of the two-blade rotor.

The quadcopter $T_q$ includes the dominant harmonic order of $n = 2$ (see blue data in Fig. \ref{fig:component_thrust_FFT}). 
The harmonic order $n=2$ of $T_q$ corresponds to the 2/rev major thrust variation, resulting from the minimum $T_q$ around $\Psi = 90\degree$ and the maximum $T_q$ around $150\degree$ during the half rotor revolution. This 2/rev variation is also strongly observed in the fuselage force $F_{q,f}$ (see yellow data in Fig. \ref{fig:component_thrust_FFT}), indicating that the download of the fuselage significantly influences the major behavior of the quadcopter thrust. 

The quadcopter rotor $4T_{q,r}$ exhibits the dominant harmonic order $n = 4$ (see red data in Fig. \ref{fig:component_thrust_FFT}), reflecting two thrust peaks around $\Psi = 90\degree$ and $150\degree$, and two thrust valleys around $\Psi = 45\degree$ and $135\degree$ (see Fig. \ref{fig:component_thrust_1rev}).
The harmonic order $n=4$ of $T_q$ is not as strong as that of $T_{q,r}$ because $4T_{q,r}$ and $F_{q,f}$ are not in phase - the phase difference is more that $(3/4)\pi$. For the harmonic order $n=6$, the phase difference between $4T_{q,r}$ and $F_{q,f}$ is nearly $\pi$, canceling almost the contribution of $T_{q,r}$ and $F_{q,f}$ to the total $T_q$. 

The quadcopter $T_q$ also includes another dominant harmonic order $n=8$, which is related to 8/rev thrust variations as shown in Fig. \ref{fig:component_thrust_1rev}. The 4/rev fluctuation of $T_{q,r}$ and the 2/rev fluctuation of $F_{q,f}$ lead to the 8/rev of $T_q$. Total 8 peaks appear in $T_q$ per revolution because (1) the two peak locations of $F_{q,f}$ are added and (2) the valley location of $F_{q,f}$ is similar to that of $T_{q,r}$ creating additional two more peaks. Therefore, the current frequency analysis reveals that the thrust variation of the quadcopter is significantly influenced by the multiple rotors and the presence of the fuselage.

\section{Conclusions \label{sec:conclusions}}

In this study, interactional aerodynamics of the quadcopter in hover was numerically investigated, focusing on major unsteady vortical structures and unsteady airloads caused by aerodynamic interaction between the quadcopter rotors and the fuselage.
The overset mesh method was employed to capture flow fields around the quadcopter using open-source flow solver OpenFOAM.
Interactional aerodynamics of the quadcopter was investigated with comparison to the isolated single rotor and relevant literature data.
Because interactional aerodynamics of multicopter causes unsteady airloads, which can eventually lead to vehicle vibration and structural fatigue, an exemplary quadcopter is chosen for the current study of interactional aerodynamics.
The major findings in this paper are as follows:

\begin{itemize}
\item Aerodynamic interaction between adjacent rotors and the fuselage can significantly influence the formation of unsteady vortical structures around the quadcopter.
\item The rotor-rotor interaction influences rotor-tip vortices to form $\Omega$-shaped vortical structures, which results in a thrust peak around the azimuthal angle $\Psi = 150\degree$ soon after the rotor-rotor interaction state of $\Psi=135 \degree$. The thrust peak is associated with the relatively large angle $\alpha$ and the incoming velocity $U_{rel}$ to the blade in the blade position $135 \degree \lesssim \Psi \lesssim 150 \degree$.
\item The fuselage disrupts the convection of the rotor wake and deflects the wake flow of the four rotors towards the center of the vehicle, which results in the fuselage download around $\Psi = 90\degree$. Although the rotor thrust has a local peak around $\Psi = 90\degree$ due to the blockage effect, it is overwhelmed by the fuselage download, causing the valley of the overall vehicle thrust.
\item The overall thrust of the quadcopter includes the major 2/rev fluctuation which consists of mainly the thrust peak around $\Psi = 150\degree$ (related to the rotor-rotor interaction) and the thrust valley around $\Psi = 90\degree$ (related to the rotor-fuselage interaction).
\item The Fourier analysis reveals that the major 4/rev fluctuation of the rotor thrust is not the second major fluctuation of the overall vehicle thrust. Instead, the 8/rev fluctuation is the second largest fluctuation for the overall vehicle thrust. This higher harmonics is associated with the multiple rotors and the presence of the fuselage in the current quadcopter configuration.
\end{itemize}

Through the exploration of complex vortical structures and unsteady airloads of the quadcopter, this study can provide insights into interactional aerodynamics of multirotor aircraft.
An in-depth understanding of interactional aerodynamics could help to design reliable and efficient multicopter aircraft, accounting for complex interactional aerodynamics between various vehicle components, including rotors and the fuselage.
Further detailed analysis in various flight conditions would help to expand our understanding of complex unsteady aerodynamics.
The authors expect relevant future studies in this field of multicopter aerodynamics.


\section*{Acknowledgments}
This work was supported by the National Research Foundation of Korea (NRF) grant funded by the Korea government (MSIT) (Project No. NRF-2021R1A2C1006193) and the National Supercomputing Center with supercomputing resources including technical support (Project No. KSC-2021-CRE-0153, KSC-2021-CRE-0372).

\bibliographystyle{elsarticle-num}
\bibliography{mybibfile}

\end{document}